\documentclass[12pt,reprint,tightenlines,groupedaddress,superscriptaddress,floatfix,aps]{revtex4}
\usepackage{graphicx}
\usepackage{rotating}
\usepackage{array}
\usepackage{multirow}
\usepackage{epstopdf}
\usepackage{amsmath}
\usepackage{amsfonts}
\usepackage{setspace}
\usepackage{braket}
\usepackage{color}

\usepackage{nameref,zref-xr} 

\usepackage{hyperref}

\DeclareMathOperator{\Tr}{Tr}

\begin{document}
\title{Evaluation of two-particle properties within finite-temperature self-consistent one-particle Green's function methods: theory and
application to GW and GF2
  } 
\author{Pavel Pokhilko}
\affiliation{Department  of  Chemistry,  University  of  Michigan,  Ann  Arbor,  Michigan  48109,  USA}
\author{Sergei Iskakov}
\affiliation{Department of Physics, University of Michigan, Ann Arbor, Michigan 48109, USA }
\author{Chia-Nan Yeh}
\affiliation{Department of Physics, University of Michigan, Ann Arbor, Michigan 48109, USA }
\author{Dominika Zgid}
\affiliation{Department  of  Chemistry,  University  of  Michigan,  Ann  Arbor,  Michigan  48109,  USA}
\affiliation{Department of Physics, University of Michigan, Ann Arbor, Michigan 48109, USA }

\renewcommand{\baselinestretch}{1.0}
\begin{abstract}
One-particle Green's function methods can model molecular and solid spectra 
at zero or non-zero temperatures. 
One-particle Green's functions directly provide electronic energies and one-particle properties, such as dipole moment. 
However, the evaluation of two-particle properties, such as $\braket{S^2}$ and $\braket{N^2}$ 
can be challenging, because they require a solution of the computationally expensive 
Bethe--Salpeter equation to find two-particle Green's functions. 
We demonstrate that the solution of the Bethe--Salpeter equation can be complitely avoided. 
Applying the thermodynamic Hellmann--Feynman theorem to  self-consistent one-particle Green's function methods, 
we derive expressions for two-particle density matrices in a general case and 
provide explicit expressions for GF2 and GW methods.  
Such density matrices can be decomposed into an antisymmetrized product of correlated one-electron density matrices and the two-particle electronic cumulant of the density matrix. 
Cumulant expressions reveal a deviation from ensemble representability for GW, 
explaining its known deficiencies. 
We analyze the temperature dependence of $\braket{S^2}$ and $\braket{N^2}$ for a set of small closed-shell systems. 
Interestingly, both GF2 and GW show a non-zero spin contamination and a non-zero fluctuation of the number of particles 
for closed-shell systems at the zero-temperature limit.

\end{abstract}
\maketitle

\renewcommand{\baselinestretch}{1.5}

\section{Introduction}

Green's function approaches\cite{Mahan00,Negele:Orland:book:2018,Martin:Interacting_electrons:2016} are complementary to the density functional theory\cite{Kohn:64:DFT,Kohn65_DFT,Parr:Weitao:DFT:1994} (DFT) and wave-function methods\cite{OlsenText,Szabo_ostlund}. 
Analogously to the Hohenberg--Kohn functional\cite{Kohn:64:DFT}  present in DFT that delivers the relationship between the ground-state energy and the ground-state density, in the Green's function formalism, there is a Green's function functional called the Luttinger--Ward functional $\Phi$ which defines the relationship between the grand canonical potential $\Omega$ and the interacting Green's function $G_{ij}(\omega)$.
The advantage of Green's function approaches lies in  a systematic and controlled way of building approximations to the Luttinger--Ward functional by including an increasing expansion of skeleton diagrams\cite{Luttinger60}. The self-energy which describes all correlation effects present in a system of interest can then be obtained as a functional derivative $\Sigma_{ij}=\frac{\partial \Phi}{\partial G_{ji}}$ with respect to the Green's function.

In contrast to wave-function approaches that deal with bulky wave functions, the one-electron Green's function $G(\omega)$ is a relatively compact object ($n\times n \times \omega_{max}$, where $n$ is the number of orbitals and $\omega_{max}$ is the size of the grid), 
requiring a much smaller storage than the many-body wave function that can easily contain millions of determinants even for a small molecular system. 
Moreover, Green's function approaches offer a distinct advantage of giving an easy and direct access to the experimentally measurable quantities  such as spectral functions, specific heats, optical spectra, or dielectric constants\cite{Stefanucci:vanLeeuwen:book:2013,Fetter:Walecka:2012}.  
Due to this easy experimental connection, Green's function approaches
are a standard computational language in many-body condensed matter physics with numerous applications to 
transport phenomena, superconductivity\cite{Kadanoff:superconductivity:1961}, and photoelectron spectroscopy\cite{Almbladh:photoemission:1985,Hedin:photoemission:1985,Fujikawa:photoelectron:chapter:2015}. 

While the one-particle Green's function gives access to the spectral function and therefore photoelectron spectrum, to study optical spectroscopy and propagation of a pair of particles, holes, or
of a particle and a hole, a two-particle Green's function is required. Similarly, in many experimental techniques (e.g. inelastic neutron scattering) spin-spin correlations functions or magnetic susceptibilities are quantities that are directly measurable. Their evaluation, however, requires a two-particle Green's function $G_{ijkl}(\omega, \omega')$ which is a bulky object since it depends on four orbital indices and two frequencies in general.
Commonly, in methods such as GW\cite{Hedin65,G0W0_Pickett84,G0W0_Hybertsen86,GW_Aryasetiawan98,Stan06,Koval14,scGW_Andrey09,GW100,Holm98,QPGW_Schilfgaarde}, the quantities that require  two-particle Green's functions are evaluated by solving the Bethe--Salpeter equation\cite{Bethe:Salpeter:1951,Onida02,Loos:BetheSalpeter:2020}. 
The Bethe--Salpeter equation for the two-particle linear-response function can be derived from
the Dyson equation employing the ``functional derivative technique'' 
of Schwinger\cite{Schwinger:derivative_technique:1959}, 
where a non-local, time-dependent, external potential $u(\omega, \omega')$ is added to the original Hamiltonian,
resulting in generalized Green functions. Subsequently,  functional derivatives of Green functions
with respect to the external potential $u(\omega, \omega')$ are evaluated to generate higher-particle Green functions. The potential $u$ is equated to zero at the end of the derivation.
While this route is formally necessary to obtain the Bethe--Salpeter equation, in practice frequently approximations to it are evaluated rather than the full complicated expressions\cite{Onida02}. 
This approach relies on an application of higher-order derivatives to a generating functional first 
and a subsequent application of approximations to the exact equations.  
This should be contrasted with the approaches that formulate approximations first and then
apply derivatives to the approximate grand potential or generating functional\cite{note:CPSCF}.

In this paper, we focus on the evaluation of the two-particle density matrix by  the application of the  thermodynamic Hellmann--Feynman theorem and two-particle perturbations to one-particle Green's functions.  
While here we consider only the time-independent two-particle density matrix, 
this theorem can also be used to evaluate time-dependent two-particle  quantities.
Such a technique allow us to avoid the solution of the computationally expensive Bethe-Salpeter equation while gaining access to the two-particle quantities.

We demonstrate that the self-consistent Green's function approximations lead to disconnected and connected (cumulant) 
parts of the two-particle density matrices. 
The explicit expressions for two-particle density matrices are given for self-consistent 
GF2\cite{Snijders:GF2:1990,Dahlen05,Phillips14,Rusakov16,Welden16} and GW\cite{Hedin65,G0W0_Pickett84,G0W0_Hybertsen86,GW_Aryasetiawan98,Stan06,Koval14,scGW_Andrey09,GW100,Holm98,QPGW_Schilfgaarde} approximations. Two-particle density matrices provide a valuable tool of interpretative 
 analysis for Green's function methods that allows us to connect with the wave function community.
Such an interpretation is important since despite the widespread applications and historical significance, 
some aspects of the Green's function methods are not well understood in the wave-function community.

In the wave-function language, the wave-function amplitudes can be analyzed to inform about the quality of the calculations.
Even if the wave function is too complex or when not explicitly available at all, 
one can use physical observables with known properties to assess the calculation quality.
For example, 
$\braket{S^2}$ is commonly used as a diagnostic for open-shell wave-function approaches\cite{Amos:spin_cont:1991,Schlegel:S2:94,Schlegel:spin_cont:1998,Stanton:CCSD_S2:1994,Krylov:S2:CC:2000} and density functional theory\cite{Baker:DFT:spin_cont:1993,Cremer:DFT:S2:2001} (DFT). 
In solids, a static spin-spin correlation function serves a similar purpose. 
$\braket{S^2}$ quantifies spin purity of the given wave function and detects the possible missing configurations, needed for spin completeness\cite{Lowdin:spin_proj:1964}. 
Applications of $\braket{S^2}$ exceed a simple diagnostic. 
It helps to access properties of the entire spin manifolds with very limited knowledge of its components via Wigner--Eckart theorem\cite{Pokhilko:SOC:19}.
A squared particle fluctuation $\braket{(\delta N)^2} = \braket{N^2} - \braket{N}^2$ is not a common diagnostic within the wave-function methods because of preservation of the number of particles, 
but it has received some attention in the context of 
particle-number symmetry breaking\cite{Scuseria:projected:quasip:2011}.
Within the wave-function methods, properties can be computed either as an expectation value 
or as a \emph{single} energy derivative with respect to perturbation\cite{OlsenText}. 
The derivative approach is especially useful for evaluation of electronic gradients and non-adiabatic couplings. 
The wave-function Hellmann--Feynman theorem\cite{Feynman:HeFe:39,Hellmannbook} provides a practical recipe of 
the derivative evaluation even for non-variational methods\cite{Handy:ZVEC:84}. 
The Hellmann--Feynman theorem can be applied within the density functional theory as well\cite{Heller:DFT:forces:1975,Painter:DFT:forces:1981,Parr:DFT:derivative:1985}.

In Green's function methods diagnostic tools such as ($\braket{S^2}$,  $\braket{(\delta N)^2} = \braket{N^2} - \braket{N}^2$) are not readily available at the level of one-particle Green's functions. Ordinarily, a two-particle Green's function needs to be constructed to access these operators. However, in the time-independent picture, a two-particle density matrix evaluated by us using the thermodynamic Hellmann-Feynman theorem is a sufficient to evaluate both quantities. In both GW and GF2, we analyze the obtained expressions for $\braket{S^2}$ and $\braket{(\delta N)^2}$ and evaluate their
values at finite temperature for a number of atomic systems. 
We show that the lack of correlated exchange in GW leads to the  density cumulant with unphysical permutational properties 
deteriorating the quality of the two-particle observables obtained. 
As a surprising result, we observe a non-zero spin contamination and 
particle number fluctuations for closed-shell systems described with GF2 and GW.

\section{Theory}
\subsection{Definitions}
The electronic Hamiltonian takes the following form\cite{OlsenText,note:nonorth}:
\begin{gather}
H = \sum_{pq} h_{pq} p^\dagger q + \frac{1}{2}\sum_{pqrs} \braket{pq | rs} p^\dagger q^\dagger sr,
\end{gather}
where all the indices run over spin-orbitals. 
Hereafter all the indices are assumed to be spin-orbitals, unless explicitly written otherwise.  
$h_{pq}$ and $\braket{pq | rs}$ are one- and two-electron integrals\cite{note:coulomb}:
\begin{gather}
h_{pq} = \int \phi_p^*(\mathbf{r};\sigma) \hat{H}_0 \phi_q(\mathbf{r};\sigma) d\mathbf{r}d\sigma, \\
\hat{H}_0 = \hat{T} + \hat{V}_{en} + \hat{V}_{nn}, \protect\label{eq:H0_def} \\
\braket{pq | rs} = (pr | qs) = 
\int \phi_p^*(\mathbf{r}_1;\sigma_1)\phi_q^*(\mathbf{r}_2;\sigma_2) \frac{1}{|\mathbf{r}_1-\mathbf{r}_2|} 
\phi_r(\mathbf{r}_1;\sigma_1)\phi_s(\mathbf{r}_2;\sigma_2) d\mathbf{r}_1 d\mathbf{r}_2 d\sigma_1  d\sigma_2.
\end{gather}
For our purposes, it is convenient to group all one-electron integrals into $\hat{H}_0$, 
the Hamiltonian of independent electrons. 
The kinetic energy operator, electron-nuclear attraction, and nuclear-nuclear repulsion are denoted as 
$\hat{T}$, $\hat{V}_{en}$, and $\hat{V}_{nn}$. 
Two-electron repulsion integrals can be written in physicists' (angle brackets) or in chemists' notation (round parenthesis). 
A particular choice of notation allows one to simplify the equations in certain cases. 
Note that unlike the conventional correlated wave-function-based methods, 
the two-electron integrals here are not antisymmetrized.

The imaginary time one-particle Green's function $G$, grand canonical partition function $Z$, 
and grand potential $\Omega$ are defined as\cite{Mahan00,Negele:Orland:book:2018} 
\begin{gather}
G_{pq} (\tau) = -\frac{1}{Z} \Tr \left[e^{-(\beta-\tau)(H-\mu N)} p e^{-\tau(H-\mu N)} q^\dagger  \right], \\
Z = \Tr \left[ e^{-\beta(\hat{H}-\mu \hat{N})} \right], \\
\Omega = -\beta \ln Z,
\end{gather}
where $\mu$ is a chemical potential and $\beta = \frac{1}{kT}$ is the inverse temperature.

Thermodynamic properties at equilibrium can be found by a thermal average defined as
\begin{gather}
\braket{O} = \frac{1}{Z} \Tr \left[ e^{-\beta(\hat{H}-\mu \hat{N})} \hat{O} \right].
\end{gather}
An expectation value of an $n$-electron operator $\hat{A}$ can be computed as 
a trace with the corresponding $n$-electron density matrix, for example,
\begin{gather}
\braket{\sum_{pq} A_{pq} p^\dagger q} = \sum_{pq} A_{pq} \braket{p^\dagger q} = \sum_{pq} A_{pq} \gamma_{pq}, \\
\braket{\sum_{pqrs} A_{\braket{pq|rs}} p^\dagger q^\dagger s r} 
= \sum_{pqrs} A_{\braket{pq|rs}} \braket{p^\dagger q^\dagger s r} 
= \sum_{pqrs} A_{\braket{pq|rs}} \Gamma_{\braket{pq|rs}}, 
\end{gather}
where $\gamma$ and $\Gamma$ are one- and two-particle density matrices.

Alternatively, thermodynamic properties can be evaluated by 
introducing a perturbation in the Hamiltonian and differentiating  with respect to the coupling strength
\begin{gather}
\hat{H}_\lambda = \hat{H} + \lambda \hat{O}, \\
\braket{O} = \frac{d \Omega}{d \lambda} = \frac{d}{d\lambda} 
\left(-\beta\ln \Tr \left[ e^{-\beta(\hat{H}-\mu \hat{N} + \lambda \hat{O})} \right]\right).
\end{gather}
The relation between these approaches is given by the thermodynamic Hellmann--Feynman theorem, 
which guarantees their equivalence for any observable $\hat{O}$ in the exact case\cite{Fan:Hellmann-Feynman:1995,Ray:Hellman-Feynman:2007,Fernandez:HellmanFeynman:2020}.
We show this equivalence for a particular case of independent electrons in Appendix A.
However, as in the case of the wave-function Hellmann--Feynman theorem, 
the equivalence may not hold for approximate methods, which we will investigate in the next section\cite{note:ExpVal}. 
Throughout the whole paper, we consider only the perturbations that do not change the atomic orbitals. 

\subsection{Implications of thermodynamic Hellmann--Feynman theorem}
A one-particle Green's function contains both static and dynamic information.
This allows one to write the grand potential as a functional of a one-particle Green's function. 
Under a perturbation with the coupling $\lambda$, the full derivative can be written as\cite{Leschke:weakSC:1976}
\begin{gather}
\frac{d\Omega[G_\lambda; \lambda]}{d\lambda} = 
\left(\frac{\partial \Omega[G_\lambda;\lambda]}{\partial \lambda}\right)_{G_\lambda} 
+ \int \frac{\delta \Omega[G_\lambda;\lambda]}{\delta G_\lambda} \frac{d G_\lambda}{d \lambda} d\mathbf{r} d\sigma d\tau,
\protect\label{eq:domdl}
\end{gather}
where $G_\lambda$ is a one-particle Green's function of the perturbed system. 
Hereafter, we follow the thermodynamic notation for partial derivatives, 
e.g., the partial derivative in the first term on the right hand side in the Eq.~(\ref{eq:domdl}) 
keeps the $G_\lambda$ constant and the differentiation is taken only with respect to the explicit dependence on $\lambda$.
If a method satisfies stationary of a grand potential with respect to the Green's function, 
only the first term survives, giving the thermodynamic Hellmann--Feynman theorem for Green's functions
\begin{gather}
\frac{\delta \Omega[G_\lambda;\lambda]}{\delta G_\lambda} = 0 \text{ for all } \lambda \Rightarrow
\frac{d\Omega[G_\lambda; \lambda]}{d\lambda} = 
\left(\frac{\partial \Omega[G_\lambda;\lambda]}{\partial \lambda}\right)_{G_\lambda}. 
\protect\label{eq:tH-F}
\end{gather}
Similar ideas have been used in the proof of the virial theorem for conserving approximations\cite{Dahlen05} and 
in the proposal for electronic gradient within Green's function methods\cite{Potthoff:density_matrix:2013}.
Self-consistent methods satisfy the Dyson equation
\begin{gather}
G^{-1} = G^{-1}_0 - \Sigma[G], \protect\label{eq:Dyson}\\
{G}^{-1}_0(i\omega_n) = i\omega_n +\mu \hat{N} - \hat{H}_0,
\end{gather}
where $G_0$ is a one-particle Green's function of independent electrons 
(note that its matrix form in non-orthogonal orbitals is given in the Eq.~\ref{eq:G_0_matrix}), 
$H_0$ is constructed according to Eq.~\ref{eq:H0_def}, and $\Sigma$ is the self-energy. 
A particular choice of the dependence of $\Sigma[G]$ determines the approximation. 

If the Dyson equation (\ref{eq:Dyson}) is satisfied, 
the explicit form of the $\Omega[G]$ is given by the Luttinger--Ward expression\cite{Luttinger60}
\begin{gather}
\Omega[G] = \Phi[G]
- \frac{1}{\beta} \sum_{\omega_m} \Tr \Sigma(i\omega_m) G(i\omega_m) 
-\frac{1}{\beta} \sum_{\omega_m} \Tr \ln(1 - G_0(i\omega_m)\Sigma(i\omega_m)) + \Omega_0, 
\protect\label{eq:LW}
\end{gather}
where $\omega_m = \frac{2\pi (2n+1)}{\beta}$ are fermionic Matsubara frequencies, 
$\Omega_0$ is the grand potential of a system of independent electrons, 
$\Phi$ is the Luttinger--Ward functional, which can be constructed perturbatively as
\begin{gather}
\Phi[G] = \sum_{n=1}^{\infty} \frac{1}{2n} \frac{1}{\beta} \sum_{\omega_m} \Tr G\Sigma^{(n)},
\protect\label{eq:Phi_LW}
\end{gather}
where $\Sigma^{(n)}$ is a perturbative contribution to the  self-energy of the order $n$. 
A direct differentiation of Eq.~(\ref{eq:Phi_LW}) gives a relation, 
defining conserving approximations\cite{Baym61,Baym62}: 
\begin{gather}
\frac{\delta \Phi}{\delta G} = \Sigma.
\end{gather}
With the Dyson equation (\ref{eq:Dyson}) this results in a stationarity of the grand potential
\begin{gather}
\frac{\delta \Omega}{\delta G} = \frac{\delta \Phi}{\delta G} - \Sigma = 0 
\end{gather}

\subsection{One-particle perturbations}
If one considers only one- and two-particle perturbations at self-consistency, 
the perturbation can be included in the integrals and
the Eq.~\ref{eq:tH-F} can be written as
\begin{gather}
\frac{d}{d \lambda}\Omega[G_\lambda, v(\lambda), h(\lambda)] = 
\sum_{pq} \left(\frac{\partial \Omega}{\partial h_{pq}}\right)_{G, v}
 \left(\frac{\partial h_{pq}}{\partial \lambda}\right)_{G, v} +
\sum_{pqrs} \left(\frac{\partial \Omega}{\partial \braket{pq|rs}}\right)_{G, h} 
\left(\frac{\partial \braket{pq|rs}}{\partial \lambda}\right)_{G, h}. 
\end{gather}
Here $h$ and $v$ denote a set of one- and two-electron integrals at some value of perturbation $\lambda$. 
For differentiation purposes, the Dyson equation can be used to rewrite Eq.~(\ref{eq:LW}) \cite{Martin:Interacting_electrons:2016} as
\begin{gather}
\Omega[G,h,v] =  \Phi[G, v]
- \frac{1}{\beta} \sum_{\omega_m} \Tr \Sigma G 
-\frac{1}{\beta} \sum_{\omega_m} \Tr \ln(-G^{-1}).  
\protect\label{eq:Omega_G}
\end{gather}
When one-electron perturbations are considered
\begin{gather}
H_0(\lambda) = H_0 + \lambda O, \\
\frac{d}{d \lambda}\Omega[G_\lambda, v, h(\lambda)] = 
- \frac{1}{\beta}  
 \sum_{\omega_m} \Tr \frac{\partial\Sigma}{\partial \lambda} G. 
\protect\label{eq:opdm_Sigma_deriv}
\end{gather}
From the Dyson equation,
\begin{gather}
\left(\frac{\partial \Sigma}{\partial \lambda}\right)_{G,v} = \frac{\partial}{\partial \lambda} G^{-1}_0.
\end{gather}
The derivative of the inverse Green's function of 
independent particles is shown in Appendix A in Eq.~(\ref{eq:der_g0_inv}).
Thus, the final expression is
\begin{gather}
\frac{d}{d \lambda}\Omega[G_\lambda, v, h(\lambda)] = 
\frac{1}{\beta} \sum_{\omega_n} \Tr O G(i\omega_n) = 
\Tr O G(0^-), 
\end{gather}
where the last equality is written in the imaginary time. 
The value of a Green's function at zero time, $G(0^-)$, 
is an expectation-value of the one-particle density matrix\cite{Martin:Interacting_electrons:2016,Mahan00,Negele:Orland:book:2018} if ensemble representability is assumed.

\subsection{Two-particle perturbations}
To evaluate two-particle properties, we include two-particle perturbations in the two-electron part of the Hamiltonian. 
If we consider perturbation of a single quadruplet of indices $p_0,q_0,r_0,s_0$, 
this leads to a value of the two-particle density matrix at this excitation
\begin{gather}
H(\lambda) = H_0 + V +
\lambda p_0^\dagger q_0^\dagger s_0 r_0 = 
H_0 + V(\lambda) \\
V(\lambda) = \frac{1}{2} \sum_{pqrs} \braket{pq|rs} p^\dagger q^\dagger s r + 
\lambda p_0^\dagger q_0^\dagger s_0 r_0= \\
\frac{1}{2} \sum_{pqrs} \left(\braket{pq|rs} + 2\lambda\delta_{p,p_0}\delta_{q,q_0}\delta_{r,r_0}\delta_{s,s_0}\right) p^\dagger q^\dagger s r = 
\frac{1}{2} \sum_{pqrs} \braket{pq|rs}_\lambda p^\dagger q^\dagger s r \\
\Gamma_{\braket{p_0 q_0|r_0 s_0}} = \frac{d\Omega}{d\lambda},
\end{gather}
where $\Gamma$ is a two-particle density matrix. 
Its index represents the notation used for integrals. 
Since not all Green's function approximations are ensemble representable, 
it is convenient to define the two-particle density matrix through this derivative rather than through an expectation value, 
since the expectation value is not defined if there is no ensemble density operator.

From the Dyson equation~\ref{eq:Dyson}, the derivative of the self-energy is
\begin{gather}
\left(\frac{\partial \Sigma}{\partial \lambda}\right)_{G,h} = 0.
\end{gather}
Therefore, the differentiation of Eq.~(\ref{eq:Omega_G}) gives
\begin{gather}
\frac{d \Omega}{d \lambda} =
\left(\frac{\partial \Omega}{\partial \lambda}\right)_{G,h} = 
\left(\frac{\partial \Phi}{\partial \lambda}\right)_{G,h} = 
\sum_{n=1}^{\infty} \frac{1}{2n} \frac{1}{\beta} \sum_{\omega_m} 
\Tr G \left(\frac{\partial\Sigma^{(n)}}{\partial \lambda}\right)_{G,h}.
\protect\label{eq:der_Phi}
\end{gather}
A two-particle density matrix, found in this way, reproduces the two-body part of the electronic energy 
when contracted with two-electron integrals. 
A detailed derivation and comparison with the Galitskii--Migdal expression is 
given in SI in section~\ref{sec:energy}.  
One can generalize Eq.\ref{eq:der_Phi} to time-dependent perturbations, introducing time into integrals. 
This approach yielding a two-particle Green's function has been used, for example, by George Baym in the context of conservation laws~\cite{Baym62} and by Robert van Leeuwen and  co-workers\cite{vanLeeuwen:xi_funct:2006} for the construction of the  $\Xi$ functional.

The post-Hartee--Fock approximations to the self-energy separate it 
into a static Hartee--Fock part and a dynamic correlated part. 
The corresponding Luttinger--Ward functional also separates into the Hartree--Fock and dynamic parts, 
providing a separation of the two-particle density matrix 
\begin{gather}
\Sigma[G] = \Sigma^\text{HF}[G] + \Sigma^\text{corr}[G], \\
\Phi[G] = \Phi^\text{HF}[G] + \Phi^\text{corr}[G], \\
\Gamma_{\braket{p_0 q_0|r_0 s_0}} = 
\Gamma_{\braket{p_0 q_0|r_0 s_0}}^\text{HF}[G] + 
\Gamma_{\braket{p_0 q_0|r_0 s_0}}^\text{corr}[G]. 
\protect\label{eq:tpdm_sep}
\end{gather}
Here all the terms are computed from the \emph{full} one-particle Green's function. 
The Hartree--Fock part of the Luttinger--Ward functional is
\begin{gather}
\Phi^\text{HF}[\gamma] = \frac{1}{2} \sum_{pqrs} \gamma_{pr} \left(\braket{pq|rs} - \braket{pq|sr}\right) \gamma_{qs},
\end{gather}
where $\gamma$ is the full correlated one-particle density matrix. 
The corresponding contribution to the two-particle density matrix is evaluated as
\begin{gather}
\Gamma_{\braket{p_0 q_0|r_0 s_0}}^\text{HF}[\gamma] = 
\gamma_{pr} \gamma_{qs} - \gamma_{ps} \gamma_{qr}.
\protect\label{eq:Gamma_HF}
\end{gather}
The antisymmetrized direct product of one-particle correlated density matrices $\Gamma^\text{HF}[\gamma]$ 
is also known as an exterior product, or a wedge product, of $\gamma$ \cite{Suhubi:exterior:2013}. 
This is a \emph{disconnected} part of the two-particle density matrix. 
Such constructions naturally occur in Green's function\cite{Negele:Orland:book:2018} and 
density matrix approaches\cite{Mazziotti:Schwinger:1998,Mazziotti:3-5-CSE:1998,Nakatsuji:RDM:1996}, 
based on Grassmann variables.
Therefore, $\Gamma^\text{corr}$ from the Eq.~(\ref{eq:tpdm_sep}) 
is the \emph{cumulant} of the two-particle density matrix\cite{Kubo:cumulant:1962}. 
This is a \emph{connected} part of the two-particle density matrix.

Below we apply Eq.~(\ref{eq:der_Phi}) to post-Hartree--Fock approximations of $\Sigma$ and analyze the numerical results. 

\subsection{GF2}
\protect\label{sec:GF2}
\begin{figure}[!h]
  \includegraphics[width=8cm]{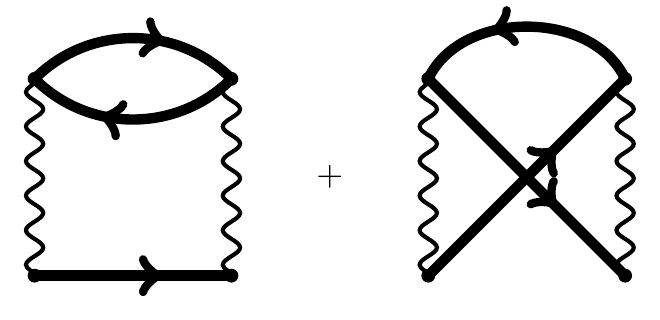}
\centering
\caption{GF2 post-HF self-energy diagrams. The diagram on the left is the correlated ``direct'' term; 
         the diagram on the right is the correlated ``exchange'' term. 
         The algebraic expressions are given in the Eqs.~\ref{eq:GF2:Sigma_direct},\ref{eq:GF2:Sigma_exchange}.
         \protect\label{fig:GF2_diag}
}
\end{figure}
The GF2 method is the second-order perturbative approximation to the  self-energy\cite{Snijders:GF2:1990,Dahlen05,Phillips14,Rusakov16,Welden16}. 
It is correlating the ``direct'' and ``exchange'' contributions:
\begin{gather}
\Sigma^{(2)} = \Sigma^\text{dir} + \Sigma^\text{ex}, \\
\Sigma_{tr}^{dir}(\tau)  = - \sum_{pqs uvw}\braket{pq | rs} \braket{tu | vw} G_{vp}(\tau) G_{wq} (\tau) G_{su} (-\tau), \protect\label{eq:GF2:Sigma_direct} \\
\Sigma_{tr}^{ex}(\tau)  = \sum_{pqs uvw}\braket{pq | rs} \braket{tu | vw} G_{wp}(\tau) G_{vq} (\tau) G_{su} (-\tau). \protect\label{eq:GF2:Sigma_exchange}
\end{gather}
For evaluation of the two-particle density matrix, 
we introduce the following intermediates
\begin{gather}
I^{dir,1}_{p_0 q_0 t s_0}(\tau) = 
- \sum_{uvw} \braket{tu | vw} G_{vp_0}(\tau) G_{wq_0} (\tau) G_{s_0u} (-\tau), \protect\label{eq:Idir1} \\
I^{dir,2}_{r q_0 r_0 s_0}(\tau) = 
- \sum_{pqs}\braket{pq | rs} G_{r_0 p}(\tau) G_{s_0 q} (\tau) G_{s q_0} (-\tau), \protect\label{eq:Idir2}\\
I^{ex,1}_{p_0 q_0 t s_0 }(\tau) = 
\sum_{uvw} \braket{tu | vw} 
G_{wp_0}(\tau) G_{vq_0} (\tau) G_{s_0u} (-\tau),  \protect\label{eq:Iex1}\\
I^{ex, 2}_{r q_0 r_0 s_0}(\tau) = 
\sum_{pqs}\braket{pq | rs}
G_{s_0 p}(\tau) G_{r_0 q} (\tau) G_{sq_0} (-\tau).  \protect\label{eq:Iex2}
\end{gather}
These intermediates are particularly convenient for use within the resolution of the identity (RI) approximation of the two-electron integrals. 
The derivatives of the self-energy are expressed through these intermediates, 
giving the final expression for the GF2 cumulant of the two-particle density matrix as
\begin{gather}
\Gamma_{\braket{p_0 q_0 | r_0 s_0}}^\text{GF2} = 
\frac{1}{4} \frac{1}{\beta}\sum_{\omega_n} \Tr 
\left(\frac{\partial \Sigma^{(2)} (i\omega_n)}{\partial \lambda}\right)_{G,h} G(i\omega_n)
= \\
\frac{1}{2}\frac{1}{\beta}\sum_{\omega_n} \Big[
\sum_{t} ( 
I^{dir,1}_{p_0 q_0 t s_0}(i\omega_n) G_{tr_0}(i\omega_n) + 
I^{ex,1}_{p_0 q_0 t s_0 }(i\omega_n) G_{tr_0}(i\omega_n) 
) +   \nonumber \\
\sum_{r} ( 
I^{dir,2}_{r q_0 r_0 s_0}(i\omega_n) G_{p_0 r}(i\omega_n) + 
I^{ex,2}_{r q_0 r_0 s_0}(i\omega_n) G_{p_0 r}(i\omega_n)
)\Big] = \\
\frac{1}{\beta}\sum_{\omega_n} \
\sum_{t} ( 
I^{dir,1}_{p_0 q_0 t s_0}(i\omega_n) G_{tr_0}(i\omega_n) + 
I^{ex,1}_{p_0 q_0 t s_0 }(i\omega_n) G_{tr_0}(i\omega_n) ). 
\end{gather}
The last equality comes from the equivalence of the terms, labeled by ``1'' and ``2''.
The spin-integrated expressions for the intermediates and the two-particle cumulant are given in SI in  Section~\ref{sec:GF2_cum}.
The final expressions can be understood through the renormalized 4-point vertex functions $\tilde{\Gamma}$, 
defined as\cite{vanLeeuwen:xi_funct:2006}
\begin{gather}
(G_2)_{ijkl} = G_{il}G_{jk} - G_{ik}G_{jl} - \sum_{pqrs} G_{ip}G_{jq}\tilde{\Gamma}_{pqrs} G_{rl} G_{sk},
\protect\label{eq:4p_vertex_def}
\end{gather}
where $G_2$ is a two-particle Green's function and its indices are written consistently with the definition below:
\begin{gather}
(G_2)_{ijkl} = \braket{T[\hat{i}(\tau_i)\hat{j}(\tau_j)\hat{k}^\dagger(\tau_k)\hat{l}^\dagger(\tau_l)]}.
\end{gather}
A careful examination of the GF2 equations (see Eq.~\ref{eq:GF2_cum_antisym}) concludes that the GF2 4-point vertex function is just an antisymmetrized two-electron integral.

A detailed numerical algorithm is shown in Appendix C.

\subsection{GW}
\protect\label{sec:GW}
The GW approximation\cite{Hedin65,G0W0_Pickett84,G0W0_Hybertsen86,GW_Aryasetiawan98,Stan06,Koval14,scGW_Andrey09,GW100,Holm98,QPGW_Schilfgaarde}  
in spin-orbitals has the following form
\begin{gather}
\Sigma_{pq}(\omega_n) = - \frac{1}{\beta}\sum_m \sum_{rs} G_{rs} (\omega_n + \Omega_m) 
\tilde{W}_{(pr|sq)} (\Omega_m), \\
W_{(i_1 i_2 | i_3 i_4)}(\Omega_n) = (i_1 i_2 | i_3 i_4) +\tilde{W}_{(i_1 i_2 | i_3 i_4)}(\Omega_n), \\
\tilde{W}_{(i_1 i_2 | i_3 i_4)}(\Omega_n) =  \sum_{i_5 i_6 i_7 i_8} 
(i_1 i_2| i_5 i_6) \Pi_{i_5 i_6 i_7 i_8} (\Omega_n) W_{(i_7 i_8 | i_3 i_4)} (\Omega_n), \\
\Pi_{i_1 i_2 i_3 i_4}(\Omega_m) = \frac{1}{\beta}\sum_{n} G_{i_2 i_3}(\omega_n) G_{i_4 i_1}(\omega_n+\Omega_m),
\end{gather}
where $\Omega_m = \frac{2\pi (2m)}{\beta}$ are bosonic Matsubara frequencies, 
$\Pi$ is the polarization function, 
$W$ is screened interaction. 
The chemical notation is convenient for writing equations in a compact way using matrix multiplications
\begin{gather}
\tilde{\mathbf{W}} = \left( \mathbf{v} \mathbf{\Pi} \mathbf{v} + \mathbf{v\Pi} \tilde{\mathbf{W}} \right), \protect\label{Wt_in_sporb} \\
\tilde{\mathbf{W}} = (1-\mathbf{v\Pi})^{-1} \mathbf{v\Pi v} = 
\left((1-\mathbf{v\Pi})^{-1}-1\right) \mathbf{v}, \\
\mathbf{W} = (1-\mathbf{v\Pi})^{-1} \mathbf{v},
\end{gather}
where the matrices in the bold font are formed by joining the neighboring spin-orbital indices into a single superindex as $(ij|kl) \rightarrow (I|K)$. 
The corresponding dynamic part of the Luttinger--Ward functional\cite{Albladh99} is 
\begin{gather}
\tilde{\Phi}^\text{GW} = -\sum_{n=1}^{+\infty} \frac{1}{2(n+1)} \frac{1}{\beta^2}\sum_{m,m^\prime}\sum_{klpq} G_{lk}(\omega_{m^\prime} + \Omega_m) \bigg[ (\mathbf{v\Pi})^n \mathbf{v} \bigg]_{(pl | kq)} G_{qp}(\omega_{m^\prime}) = \\
-\frac{1}{2} \frac{1}{\beta} \sum_{\Omega_m} \Tr \sum_{n=2}^\infty \frac{(\mathbf{v\Pi})^n}{n} = 
-\frac{1}{2} \frac{1}{\beta} \sum_{\Omega_m} \Tr \left( \sum_{n=1}^\infty \frac{(\mathbf{v\Pi})^n}{n} -\mathbf{v\Pi}\right) = \\
\frac{1}{2} \frac{1}{\beta} \sum_{\Omega_m} \Tr \left( \ln(1-\mathbf{v\Pi}) +\mathbf{v\Pi}\right). 
\protect\label{eq:GW_Phi}
\end{gather}
It is possible to use an alternative functional $\Psi[G, W]$, defined as the Legendre transform of $\Phi[G, v]$:
\begin{gather}
\Psi[G, W] = \Phi[G, v[G, W]] -\frac{1}{2}\frac{1}{\beta} \sum_{\Omega_m}\Tr \big[ \mathbf{\Pi W} - \ln(1+\mathbf{\Pi W}) \big].
\end{gather}

Trace operations allow one to take derivatives of matrix functions in the same way as it is done for functions of a single variable. 
Differentiating Eq.~(\ref{eq:GW_Phi}), we get
\begin{gather}
\Gamma^{\text{GW}} = \left(\frac{\partial \tilde{\Phi}^\text{GW}}{\partial \lambda}\right)_{G,h} = 
-\frac{1}{2}\frac{1}{\beta} \sum_{\Omega_m} 
\big( \Tr ((1-\mathbf{v\Pi})^{-1}-1)\frac{\partial \mathbf{v}}{\partial \lambda}\mathbf{\Pi}  \big) = \\
-\frac{1}{2} \frac{1}{\beta}\sum_{\Omega_m} 
\big( \Tr (1-\mathbf{v\Pi})^{-1}\mathbf{v\Pi}\frac{\partial \mathbf{v}}{\partial \lambda}\mathbf{\Pi}  \big) = \\
-\frac{1}{2} \frac{1}{\beta}\sum_{\Omega_m} 
\big( \Tr \mathbf{W\Pi}\frac{\partial \mathbf{v}}{\partial \lambda}\mathbf{\Pi}  \big). 
\end{gather}
The RI approximation, used in GW to lower the computational cost, decomposes integrals into 3-index tensors
\begin{gather}
(pq|rs) = \sum_Q V_{pq}^Q V_{rs}^Q,
\end{gather}
where $Q$ is the auxiliary index in AO, provided by an auxiliary basis set. 
$\tilde{W}$ is written as
\begin{gather}
\tilde{W}_{pqrs}(\Omega_n) = \sum_{Q,Q^\prime} V_{pq}^Q \tilde{P}_{QQ^\prime}(\Omega_n) V_{rs}^{Q^\prime}, \\
\tilde{P}(\Omega_n) = \left(1-\tilde{P}_0(\Omega_n)\right)^{-1} \tilde{P}_0(\Omega_n), \\
\tilde{P}_{0,QQ^\prime}(\Omega_n) =  -\sum_{m} \sum_{pqrs} V_{pq}^Q G_{ps}(\omega_m) G_{rq}(\omega_m+\Omega_n)V_{rs}^Q,
\end{gather}
where $\tilde{P}$ is a renormalized polarization matrix\cite{Iskakov20}. 
This leads to the following expression for the two-particle density matrix cumulant
\begin{gather}
\Gamma_{(p_0 q_0 | r_0 s_0)}^{\text{GW}} = 
-\frac{1}{\beta}\sum_{\Omega_m} 
\sum_{pqrs} \Pi_{r_0 s_0 pq}(\Omega_m) W_{(pq|rs)}(\Omega_m) \Pi_{rs p_0 q_0}(\Omega_m) = \\
-\frac{1}{\beta}\sum_{\Omega_m} 
\sum_{pqrs} \Pi_{r_0 s_0 pq}(\Omega_m) V_{pq}^Q (\delta_{Q,Q^\prime}+\tilde{P}_{QQ^\prime}(\Omega_m))  V^{Q^\prime}_{rs}\Pi_{rs p_0 q_0}(\Omega_m). 
\protect\label{eq:GW_cumulant}
\end{gather}
This equation and Eq.~\ref{eq:4p_vertex_def} also give the GW renormalized 4-point vertex, which is $W$.
The numerical algorithm for an evaluation of the GW cumulant is given in Appendix D.

\section{Results and discussion}
\protect\label{sec:results}

\subsection{Computational details}
We investigated the temperature dependence of two-particle properties, computed with GF2 and GW. 
We applied this formalism to a set of closed-shell systems: 
\begin{enumerate}
\item Noble gases: He, Ne, Ar atoms.
\item Alkaline earth metals: Be, Mg, Ca atoms.
\end{enumerate}
We used Dunning's correlation consistent 
double-zeta cc-pVDZ basis sets~\cite{Dunning:ccpvxz:He,Dunning:ccpvxz:LiNaBeMg,Dunning:ccpvxz:1989,Dunning:ccpvxz:Al-Ar,Dunning:ccpvxz:Ca}, 
taken from the EMSL Basis Set Exchange website~\cite{NewBSE,EMSL-paper}. 
All electrons were correlated.
We used an intermediate representation\cite{Yoshimi:IR:2017} for the grid with the $\Lambda = 10^5$ and 136 functions. 
$10^{-8}$~a.u. threshold was used as a convergence criterion for the energy.
An RI approximation was used in all calculations. 
Integrals and even-tempered auxiliary RI basis sets were generated by the PySCF program~\cite{PYSCF}.
To perform the GW and GF2 calculations, we used the local in-house code for solids, used previously for NiO and MnO\cite{Iskakov20} solids.

\subsection{Low temperature}
\begin{table} [tbh!]
  \caption{$\braket{S^2}$ and $(\delta N)^2$ at $\beta = 1000$~a.u.$^{-1}$ ($36.7$~eV$^{-1}$), 
           computed from full ($\Gamma^{\text{full}}$) and 
           disconnected parts ($\Gamma^{\text{HF}}[\gamma]$) of the two-particle density matrix. 
\protect\label{tbl:S2_N2}}
\begin{tabular}{l|cc|cc}
\hline
GF2        &  \multicolumn{2}{c|}{$\braket{S^2}$}   &  \multicolumn{2}{c}{$(\delta N)^2$}     \\       
System     &  $\Gamma^\text{full}$ & $\Gamma^\text{HF}[\gamma]$ &
              $\Gamma^\text{full}$ & $\Gamma^\text{HF}[\gamma]$ \\
\hline
He         & 0.0133 & 0.0133 & 0.0177 & 0.0177 \\ 
Ne         & 0.0768 & 0.0767 & 0.1011 & 0.1022 \\ 
Ar         & 0.1233 & 0.1230 & 0.1605 & 0.1640 \\ 
Be         & 0.0819 & 0.0812 & 0.1048 & 0.1083 \\ 
Mg         & 0.0796 & 0.0790 & 0.1025 & 0.1054 \\ 
Ca         & 0.1841 & 0.1829 & 0.2327 & 0.2438 \\ 
\hline
GW         &  \multicolumn{2}{c|}{$\braket{S^2}$}   &  \multicolumn{2}{c}{$(\delta N)^2$}     \\       
System     &  $\Gamma^\text{full}$ & $\Gamma^\text{HF}[\gamma]$ &
              $\Gamma^\text{full}$ & $\Gamma^\text{HF}[\gamma]$ \\
\hline
He         & 0.3538 & 0.0200 & 0.4716  & 0.0267  \\
Ne         & 1.0536 & 0.0663 & 1.4043  & 0.0884  \\
Ar         & 1.5735 & 0.1110 & 2.0963  & 0.1480  \\
Be         & 1.1016 & 0.0941 & 1.4658  & 0.1255  \\
Mg         & 1.1863 & 0.0933 & 1.5791  & 0.1244  \\
Ca         & 2.2502 & 0.1725 & 2.9940  & 0.2301  
\end{tabular}
\end{table}
Table~\ref{tbl:S2_N2} shows the computed values of $\braket{S^2}$ (a.u.) and squared particle fluctuations $(\delta N)^2 = \braket{N^2}-\braket{N}^2$ at low temperature. 
Their computational expressions are given in the Section~\ref{sec:S2_N2} in the SI, 
which are not the same as for the wave-function methods\cite{Stanton:CCSD_S2:1994}, 
because the fluctuating number of electrons.
Both GF2 and GW show spin contamination of the closed-shell ground states 
and non-zero fluctuation of the number of particles even for a very low temperature. 
According to NIST atomic data~\cite{NIST:5.8:2020},
the lowest excited states are more than 1.8~eV above the ground state. 
Thus, the residual spin contamination at $\beta = 1000$~a.u.$^{-1}$ ($36.7$~eV$^{-1}$). 
should be by many orders of magnitude smaller than the ones reported in Table~\ref{tbl:S2_N2}. 
Our observation is consistent with emerging evidence of spin contamination in closed-shell molecules 
for perturbative methods, such as MP2 and CC2\cite{Stopkowicz:closed_shell:spin_cont:2021}.

The values of $\braket{S^2}$ and $\braket{(\delta N)^2}$ are correlated with each other. 
This can be rationalized as follows. 
If one neglects the cumulant, the fluctuation of any property is given by the exchange contribution to the $\Gamma^{\text{HF}}[\gamma]$. 
As clear from the spin-integrated expressions (eq.~\ref{eq:Gamma_HF_aaaa}--\ref{eq:Gamma_HF_bbbb} in SI), 
the only non-zero contributions to the fluctuation come from the same-spin parts. 
From all the expressions for one- and two-particle density matrices, $\braket{S_-S_-} = \braket{S_+S_+} = 0$. 
Therefore, for zero spin projection $\braket{S_z} = 0$, $\braket{S_x^2} = \braket{S_y^2} = \braket{S_z^2} = \frac{1}{3}\braket{S^2}$.
Thus, comparing the Eq.~\ref{eq:S2} and \ref{eq:N2} from SI and dropping the opposite-spin parts, we get
\begin{gather}
\Tr \big[ \Gamma^{\text{HF}}[\gamma] S^2 \big] = \frac{3}{4} \Tr \big[ \Gamma^{\text{HF}}[\gamma] N^2 \big].
\protect\label{eq:N2_S2_relation}
\end{gather}
Here and everywhere else in the paper these disconnected contributions are evaluated with the 
full correlated one-particle density matrix $\gamma$. 
If cumulant is included, the opposite-spin contributions to fluctuations are no longer zero (because the cumulant includes description of Coulomb hole), 
and the equality from Eq.~(\ref{eq:N2_S2_relation}) does not hold for the full two-particle density matrix. 

For a weakly correlated system, the cumulant part of two-particle density matrix 
is expected to have a minor impact on two-particle properties. 
This is shown in Table~\ref{tbl:S2_N2} for the atomic closed-shell systems considered with the GF2 method. 
The GW cumulant, however, significantly worsens the spin contamination and particle fluctuations. 
The drastic difference between GF2 and GW can be explained from the symmetry properties of the obtained cumulants. 
A careful investigation of equations~\ref{eq:Idir1}--\ref{eq:Iex2} reveals that intermediates $I^{dir,1}$ and $I^{ex,1}$, $I^{dir,2}$ and $I^{ex,2}$ can be grouped together and expressed through antisymmetrized integrals:
\begin{gather}
\braket{pq || rs} = \braket{pq | rs} - \braket{pq | sr} \\
I^{dir+ex,1}_{p_0 q_0 t s_0}(\tau) = 
- \sum_{uvw} \braket{tu || vw} G_{vp_0}(\tau) G_{wq_0} (\tau) G_{s_0u} (-\tau) \protect\label{eq:I1} \\
I^{dir+ex,2}_{r q_0 r_0 s_0}(\tau) = 
- \sum_{pqs}\braket{pq || rs} G_{r_0 p}(\tau) G_{s_0 q} (\tau) G_{s q_0} (-\tau). \protect\label{eq:I2}
\end{gather}
The resulting GF2 cumulant is defined as 
\begin{gather}
\Gamma_{\braket{p_0 q_0 | r_0 s_0}}^\text{GF2} = 
\frac{1}{\beta}\sum_{\omega_n} 
\sum_{t}  
I^{dir+ex,1}_{p_0 q_0 t s_0}(i\omega_n) G_{tr_0}(i\omega_n).
\protect\label{eq:GF2_cum_antisym}
\end{gather}
Expressing the sums over Matsubara frequencies as a convolution in the imaginary time domain, 
one can see that the GF2 cumulant inherits the permutational properties of the antisymmetrized integrals. 
In the wave-function methods, permutational properties of this type are ensured starting from the Hamiltonian expressed through antisymmetrized integrals. 
Such approaches lead to specific permutational properties of the wave-function amplitudes, which can be written through time-independent antisymmetrized Goldstone diagrams\cite{BartlettShavitt:CC}. 

Marios-Petros Kitsarasa and Stella Stopkowicz observed non-zero $\braket{S^2}$ 
values for perturbative MP2 and CC2 for closed-shell molecules\cite{Stopkowicz:closed_shell:spin_cont:2021}. 
Their explanation lies in the computational expression for $\braket{S^2}$ that contains one- and two-electron parts. 
The one- and two-particle density matrices, 
evaluated through the first derivative of Lagrangians and Hellmann--Feynman theorem, 
are correct up to the second and first perturbative orders, respectively. 
This inconsistency in perturbation orders of additive contributions leads to unphysical values of $\braket{S^2}$ . 
This perturbative analysis is conceptually close to diagrammatic expansions in terms of 
non-interacting Green's function $G_0$ and \emph{bare} two-electron interaction. 
We adapt this idea and generalize it to ``bold'' perturbative expansions 
with the bare two-electron interaction and full (``bold'') Green's function. 
The expressions for one- and two-particle properties (Eq.~\ref{eq:opdm_Sigma_deriv} and Eq.~\ref{eq:der_Phi}) 
are fundomentally different. 
The differentiation of the self-energy with respect to one-particle perturbations does not change 
the perturbative order with respect to the bare interaction 
(diagrammatically, the number of interaction lines is preserved). 
However, differentiation of the self-energy over two-particle perturbations reduces  
the perturbative order with respect to the bare interaction by one
(one interaction line is removed). 
In a finite perturbative order with respect to the bare interaction, 
the self-energy and 4-point vertex functions do not have consistent perturbative orders, 
leading to inconsistent perturbative orders of one- and two-particle matrices. 
In particular, GF2 one-particle properties are correct up to the second perturbative order, 
while two-particle properties are correct only up to the first perturbative order, 
which leads to unphysical values of $\braket{S^2}$ and $\braket{(\delta N)^2}$ in the zero-temperature limit. 

In GW, the screened interaction $W$ has contributions with up to an infinite perturbative order of the bare interaction. 
Therefore, there is no inconsistency in perturbative orders over the bare interaction 
for one- and two-particle density matrices. 
Yet, GW spin contamination and particle-number fluctuations are worse than GF2 ones, 
which can be rationalized as follows. 
GW is an example of an approximation that cannot be expressed in terms of antisymmetrized integrals. 
The screened interaction $W$ inherits the symmetry of the non-antisymmetrized integrals, representing a lack of correlated exchange\cite{note:crossing}. 
This gives a cumulant with unphysical properties, explaining a deterioration of the computed observables:
\begin{gather}
\Gamma^{\text{GW}}_{\braket{pq|rs}} \neq 
-\Gamma^{\text{GW}}_{\braket{pq|sr}} \\
\Gamma^{\text{GW}}_{\braket{pq|rs}} \neq 
-\Gamma^{\text{GW}}_{\braket{qp|rs}} .
\end{gather}
This violation happens already in the first perturbative order for the two-particle density matrix, 
while the first perturbative order of the self-energy equals to its Hartree--Fock part with the correct account of exchange. 
A lack of the correlated exchange in the self-energy happens only in the second perturbative order with 
respect to the bare interaction. 
Since the inconsistency in the treatment of the exchange happens already in the first order of the two-particle density matrix, 
the GW spin contamination is worse than the one in  GF2. 
Since the antisymmetric structure is a necessary condition for an ensemble representability 
following from the anticommutation relations of fermionic creation and annihilation operators, the GW two-particle density matrix violates the ensemble representability 
(generalization of the $N$-representability for ensembles).  
Approximate variational 2-RDM methods yield 
total energies below full CI due to a violation of 
some of the $N$-representability conditions\cite{Mazziotti:vRDM:2002,Mazziotti:vRDM:2006}.
This explains why GW total energies are systematically below full CI\cite{MB_comparison:2020,Tran_GW_SEET}. 

\subsection{Temperature dependence}
\begin{figure}[!h]
  \includegraphics[width=10cm]{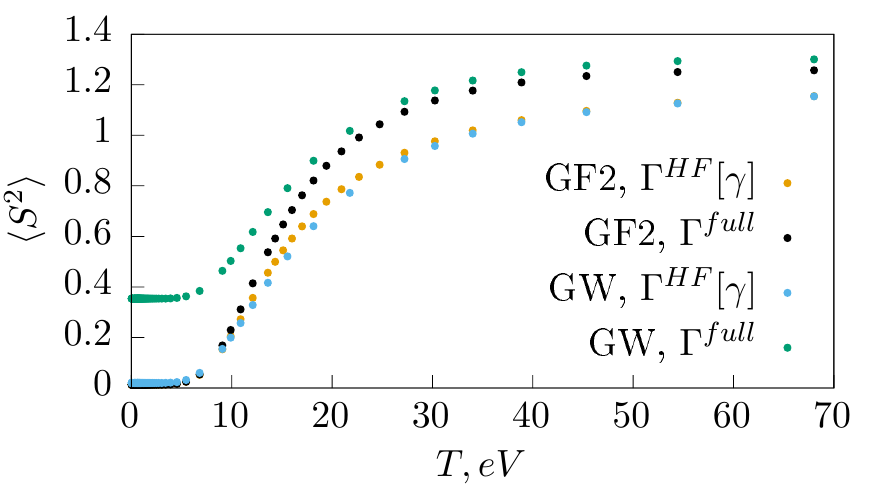}\\
  \includegraphics[width=10cm]{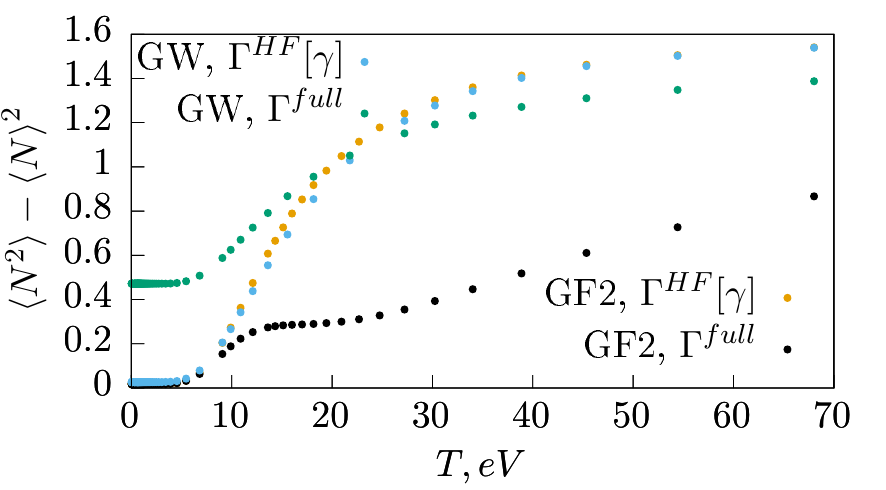}
\centering
\caption{The temperature dependence of $\braket{S^2}$ (top) and $\braket{N^2}-\braket{N}^2$ (bottom) for a helium atom. 
         Contributions of the antisymmetrized Kroneker product of one-particle density matrices (Eq.~\ref{eq:Gamma_HF}) 
         and full two-particle contributions are shown for both GF2 and GW.
         \protect\label{fig:He}}
\end{figure}
Figure~\ref{fig:He} shows the temperature dependence of $\braket{S^2}$ and $\braket{(\delta N)^2}$ 
for a helium atom. 
An increase in the temperature causes enhanced fluctuations of spin and number of particles. 
As the temperature reaches 10 eV, the total $\braket{S^2}$ and $\braket{(\delta N)^2}$ 
deviate from the values, computed with $\Gamma^{\text{HF}}[\gamma]$. 
In this temperature range, the  entanglement between electrons is effectively increased making the role of the cumulant more significant. 
The further growth of temperature gradually leads to a high-temperature limit, 
where methods converge to a HF solution. 
Neon and argon atoms show a similar behavior, shown in the Section~\ref{sec:finite_T_graphs} in SI.

\begin{figure}[!h]
  \includegraphics[width=8cm]{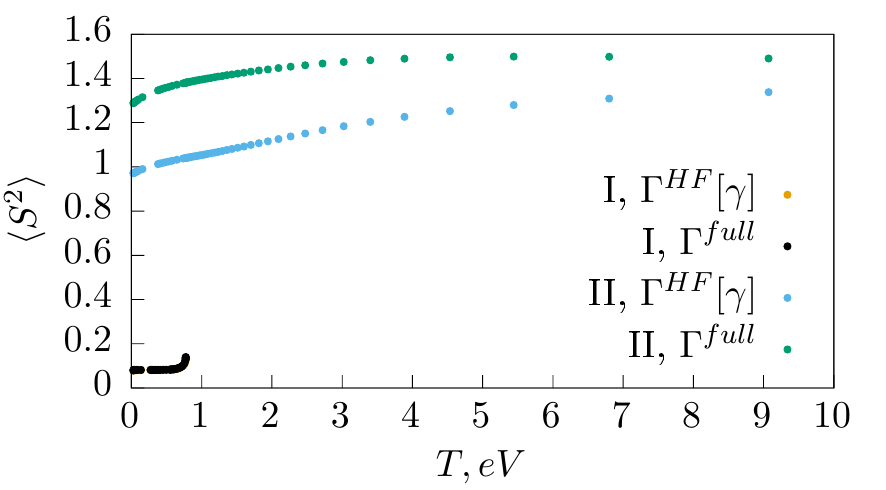} 
  \includegraphics[width=8cm]{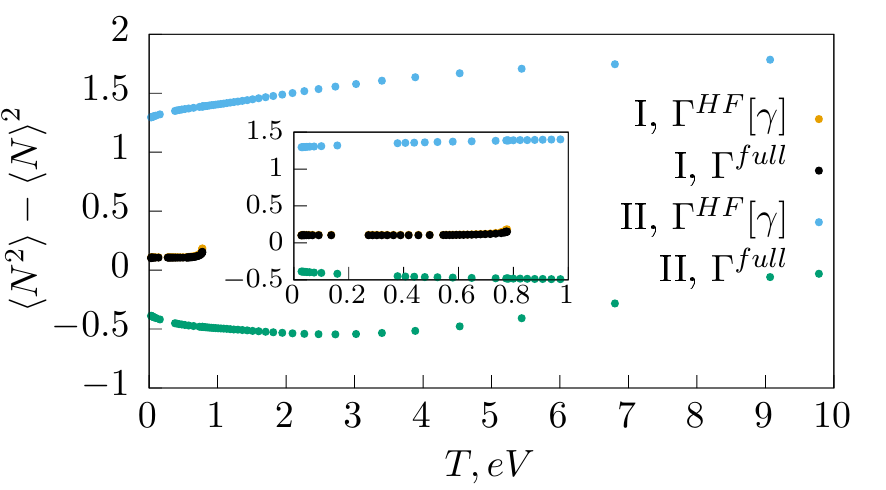} \\
  \includegraphics[width=8cm]{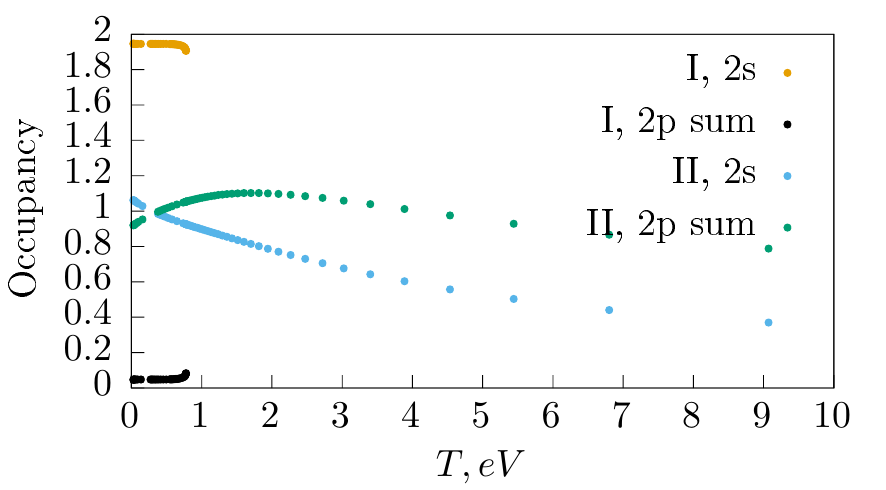} 
  \includegraphics[width=8cm]{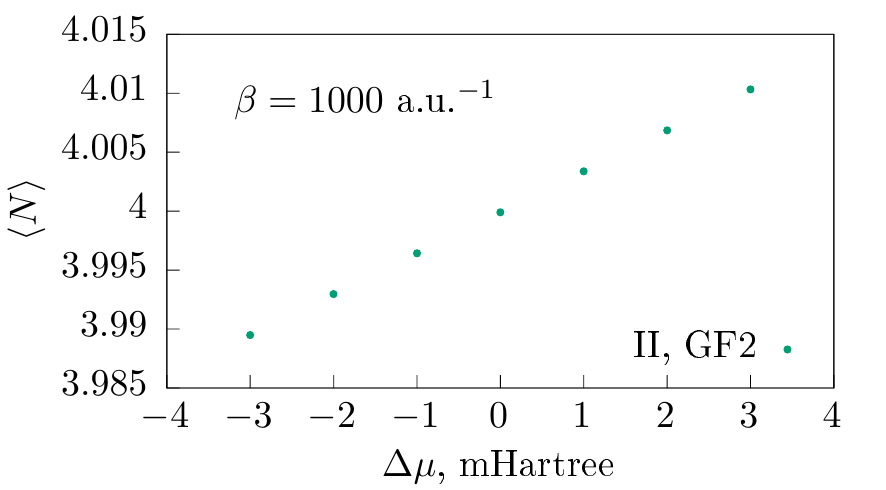}
\centering
\caption{Top: The temperature dependence of $\braket{S^2}$ (left) and $\braket{N^2}-\braket{N}^2$ (right) evaluated by GF2 for a beryllium atom. 
         Two solutions of the Dyson equation are shown. 
         The \textbf{solution I} corresponds to a closed shell at a low temperature. 
         The \textbf{solution II} is a high-temperature solution with nearly equal occupations of 2s and 2p orbitals. 
         Its low-temperature limit corresponds to a 
         broken-spin Hartree--Fock solution with fractional occupations. 
         Bottom, left: Orbital occupancies for both solutions. 
         Bottom, right: The dependence of the average number of electrons on the chemical potential.
         \protect\label{fig:Be}}
\end{figure}

Beryllium atom gives a qualitatively different picture, shown in the Figure~\ref{fig:Be}. 
The GF2 solution (denoted as \textbf{solution I}), converging to the ground state of Be atom in the low temperature limit, exists only up to the temperature of approximately 1 eV. 
Above this temperature, the calculations swapped to another solution (denoted as \textbf{solution II}) with 2s orbital occupation close to 1 and occupations of each of the 2p$_x$, 2p$_y$, and 2p$_z$ close to $1/3$. 
This solution is spin contaminated even at a low temperature, indicating spin symmetry breaking. 
There is a Hartree--Fock solution with these fractional occupancies that is close to the GF2 solution. 
The solutions of this type are sometimes referred within Slater--Hartee--Fock method that generalizes Hartree--Fock to fractional occupancies\cite{Slater:FON:HF:1969,Zigner:ensemble:HF:1977}.  
The GF2 cumulant has a strong contribution to the fluctuation of the number of particles, 
giving a negative value of $\braket{(\delta N)^2}$. 
This means that the semi-positive definiteness of the GF2 two-particle density matrix does not hold for this solution. 
Similar solutions are observed for calcium and magnesium atoms, shown in the Section~\ref{sec:finite_T_graphs} in SI. 
GW does not lead to two solutions, and the transition to high temperature is continuous.

If the description of an ensemble is exact, the fluctuation of the number of particles is connected with the derivatives over the chemical potential
\begin{gather}
\braket{N^2}-\braket{N}^2 = \frac{1}{\beta} \left(\frac{\partial\braket{N}}{\partial\mu}\right)_{T,V}
 = -\frac{1}{\beta} \left(\frac{\partial^2 \Omega}{\partial\mu^2}\right)_{T,V}.
\protect\label{eq:fluct_der}
\end{gather}
However, this equality may not hold if approximate methods are used. 
For the \textbf{solution II} of the Be atom, $\braket{(\delta N)^2} = -0.38807$ at $\beta = 1000$~a.u.$^{-1}$ ($36.7$~eV$^{-1}$). 
The dependence of the average number of electrons for this solution on the chemical potential is shown in Figure~\ref{fig:Be}. 
The numerical derivative evaluated with 7-point stencil is $\partial\braket{N}/\partial\mu = 3.47123$~a.u.$^{-1}$. 
Additional details for other stencils are shown in the Table~\ref{tbl:fdiff_n_mu} in SI. 
Thus, for GF2 the equality~\ref{eq:fluct_der} is violated strongly---different sides have different signs.

\subsection{Connection with the wave-function approaches}
Single-reference wave-function methods often start from a selected vacuum determinant. 
From the time-independent Wick's theorem, 
the two-particle density matrix decomposes into separable and non-separable parts\cite{sfccsdgrad}:
\begin{gather}
\Gamma_{\braket{pq|rs}} = \Gamma_{\braket{pq|rs}}^{\text{sep}} + \Gamma_{\braket{pq|rs}}^{\text{non-sep}}, \protect\label{eq:Gamma_sep_nonsep} \\
\Gamma_{\braket{pq|rs}}^{\text{sep}} = \rho_{rp}\gamma_{qs} -\rho_{rq}\gamma_{ps} + 
\rho_{sq}\gamma_{pr} -\rho_{sp}\gamma_{qr}, \protect\label{eq:Gamma_sep}
\end{gather}
where $\rho$ is the density of the reference determinant. 
This expression does not only simplifies the derivations of computational expressions, 
but also allows one to develop approximations for two-particle properties. 
As shown in the Ref.\cite{Krylov:EOMEESoc}, taking only separable part of transition two-particle density matrix for spin--orbit couplings (SOC) gives the spin--orbit mean-field approximation (SOMF)\cite{Marian:SOCs:2012,Hess:96:MeanField}. 
SOMF has been shown to be a very good approximation to the full SOC, 
as demonstrated in multiple publications\cite{Ruud:SOMFbench:1999,Gauss:00:SOCC,Gauss:EOMIPsoc:08,Neese:SOCMF:05,Berning:SOCs:2000,Marian:SOCs:2012,Krylov:EOMEESoc,Kaupp:GtensSOMF:02,Cheng:2comp:MF:2018,Cheng:X2C:SOMF:2018,Pokhilko:SOC:19,Coriani:Ledges:20,Casanova:RASCI_SOC:2020}. 

The Eq.~(\ref{eq:tpdm_sep}) \emph{differs} from the Eqs.~\ref{eq:Gamma_sep_nonsep} and \ref{eq:Gamma_sep}. 
In Hermitian theory, Eq.~(\ref{eq:tpdm_sep}) can be 
``derived'' from Eqs.~\ref{eq:Gamma_sep_nonsep} and \ref{eq:Gamma_sep} by taking $\rho = \gamma$. 
Such density matrix may not be representable by a single determinant wave function. 
Generalized Wick's theorem\cite{Mukherjee:Wick:1997,Kutzelnigg:Wick:1997} makes this connection rigorous. 
When normal ordering is performed with respect to the entire wave function, 
generalized Wick's theorem gives exactly the Eq.(\ref{eq:tpdm_sep}) 
as was shown for $k$-particle density matrices\cite{Kutzelnigg:cumulant:1999}. 
Application of the generalized Wick's theorem to thermal ensembles
 makes a formal basis for formulation of thermal theories\cite{Hirata:Bartlett:thermal_CCSD:2004,Hirata:Kohn-Littunger:2013,Hirata:thermal_CCSD:MBPT:2015}.

\section{Conclusions}
We presented an application of the thermodynamic Hellmann--Feynman theorem to
self-consistent one-particle Green's function methods. 
This general formalism provides arbitrary time-dependent and time-independent properties, 
including one- and two-particle properties. 
We derived the corresponding expressions up to an arbitrary perturbative order. 
Such density matrices fully reproduce the total energy computed from the Galitskii--Migdal formula. 
The explicit computational expressions for evaluation of both one- and two-particle density matrices and the corresponding numerical algorithms are given for both GF2 and GW. 
Two-particle density matrices decompose into an antisymmetrized product of the correlated one-particle density matrices 
and the electronic cumulant. 
The structure of the electronic cumulant, established within respective approximations, reveals a violation of the ensemble representability for GW. 
Numerical applications to the set of atoms illustrate usage of 
$\braket{S^2}$ and $\braket{N^2}$ quantities as diagnostic tools, 
providing an insight into the underlying electronic structure. 
In particular, we found that both GF2 and GW show a non-zero spin contamination and non-zero fluctuation of the number of particles at zero-temperature limit for closed-shell systems, 
which we explained through perturbave series over the bare interaction. 
The presented derivation and analysis are useful for evaluations of local correlators, 
such as spin and charge correlators, explaining the electronic structure of materials. 
The density matrices obtained can also be used in a framework of energy decomposition analysis, 
quantifying various contributions into the total energy.  

\section*{Acknowledgments}
P. P., Ch-N. Y., and D. Z.  were supported by the U.S. Department of Energy under Award No. DE-SC0019374.
S. I. is supported by the Simons foundation via the Simons Collaboration on the Many-Electron Problem. \\

\section*{Supplementary Material}
If non-orthogonal spin-orbitals are used, the anticommutating relations between creation and annihilation operators are
\begin{gather}
\{ p^\dagger, q^\dagger \} = 
\{ p, q \} = 0 \\
\{ p^\dagger, q \} = S^{-1}_{pq},
\end{gather}

\section*{Supplementary Material: Expressions for $\braket{S^2}$ and $(\delta N)^2$} 
\protect\label{sec:S2_N2}
All the expressions here are derived under assumption that the second-quantized anticommutation relations 
are valid when the thermal average is taked. 
This may not be the case, for example, if an approximate method is not ensemble representable, 
which will lead to numerical artifacts. 
It is also assumed that the thermal averages of excitation operators 
result in the matrix elements of the corresponding density matrices.

Spin ladder operators and $S^2$ are
\begin{gather}
S_+ = S_x + iS_y \\
S_- = S_x - iS_y \\
S^2 = S_x^2 + S_y^2 + S_z^2 = \frac{1}{2} ( S_+ S_- + S_- S_+ ) + S_z^2 = \\
\frac{1}{2} ( [S_+, S_-] + 2S_- S_+ ) + S_z^2 = \\
S_-S_+ +  S_z + S_z^2
\end{gather}
Because we work with finite-temperature methods, $\braket{S_z^2} \neq \braket{S_z}^2$.
In second quantization (in spin-orbitals), these operators are
\begin{gather}
S_\mu = \sum_{pq} \braket{p|S_\mu|q} p^\dagger q \\
S_\mu S_\nu = \sum_{pqrs} \braket{p|S_\mu|q} \braket{r|S_\nu|s} p^\dagger q r^\dagger s = \\
-\sum_{pqrs} \braket{p|S_\mu|q} \braket{r|S_\nu|s} p^\dagger r^\dagger q  s + \\
\sum_{pqrs} \braket{p|S_\mu|q}S_{qr}^{-1} \braket{r|S_\nu|s} p^\dagger  s 
\end{gather}

\begin{gather}
\braket{S_-S_+} = 
-\sum_{pqrs} \braket{p|S_-|q} \braket{r|S_+|s} \Gamma_{prsq}  + \\
\sum_{pqrs} \braket{p|S_-|q} S^{-1}_{qr}\braket{r|S_+|s} \gamma_{ps} = \\
-\sum_{pqrs} \braket{p|q} \braket{r|s} \Gamma_{prsq}^{\beta\alpha \beta \alpha}  + \\
\sum_{pqrs} \braket{p|s} \gamma_{ps}^{\beta\beta} 
\end{gather}
The last equality is written in AO.
\begin{gather}
\braket{S_+S_-} = 
-\sum_{pqrs} \braket{p|S_+|q} \braket{r|S_-|s} \Gamma_{prsq}  + \\
\sum_{pqrs} \braket{p|S_+|q} S^{-1}_{qr}\braket{r|S_-|s} \gamma_{ps} = \\
-\sum_{pqrs} \braket{p|q} \braket{r|s} \Gamma_{prsq}^{\alpha\beta \alpha \beta}  + \\
\sum_{pqrs} \braket{p|s} \gamma_{ps}^{\alpha\alpha} 
\end{gather}
\begin{gather}
S_z^2 = \frac{1}{4} \sum_{pqrs} S_{pq}S_{rs} (a^\dagger_{p\alpha}a_{q\alpha} - a^\dagger_{p\beta}a_{q\beta})
(a^\dagger_{r\alpha}a_{s\alpha} - a^\dagger_{r\beta}a_{s\beta}) = \\
\frac{1}{4} \sum_{pqrs} 
S_{pq}S_{rs} (
+a^\dagger_{p\alpha}a_{q\alpha} a^\dagger_{r\alpha}a_{s\alpha}
-a^\dagger_{p\alpha}a_{q\alpha} a^\dagger_{r\beta}a_{s\beta}
-a^\dagger_{p\beta}a_{q\beta} a^\dagger_{r\alpha}a_{s\alpha}
+a^\dagger_{p\beta}a_{q\beta} a^\dagger_{r\beta}a_{s\beta}) = \\
\frac{1}{4} \sum_{pqrs} 
S_{pq}S_{rs} (
-a^\dagger_{p\alpha} a^\dagger_{r\alpha}a_{q\alpha} a_{s\alpha}
-a^\dagger_{p\alpha} a^\dagger_{r\beta} a_{s\beta} a_{q\alpha} 
-a^\dagger_{p\beta} a^\dagger_{r\alpha} a_{s\alpha} a_{q\beta} 
-a^\dagger_{p\beta} a^\dagger_{r\beta} a_{q\beta} a_{s\beta}) + \nonumber \\
\frac{1}{4} \sum_{pqrs} S_{pq} S_{rs} S^{-1}_{qr} (
a^\dagger_{p\alpha} a_{s\alpha}
+a^\dagger_{p\beta} a_{s\beta}) \\
\braket{S_z^2} = 
-\frac{1}{4} \sum_{pqrs} \braket{p|q}\braket{r|s} (
\Gamma_{prsq}^{\alpha\alpha\alpha\alpha}+
\Gamma_{prqs}^{\alpha\beta\alpha\beta}+
\Gamma_{prqs}^{\beta\alpha\beta\alpha}+
\Gamma_{prsq}^{\beta\beta\beta\beta}) \nonumber \\
+ \frac{1}{4}\sum_{ps} \braket{p|s} (\gamma_{ps}^{\alpha\alpha} + \gamma_{ps}^{\beta\beta}) \protect\label{eq:S2}
\end{gather}
Again, the last equality here is written in AO. 
\begin{gather}
\braket{S_z} = 
\frac{1}{2}\sum_{pq} \braket{p|q} (\gamma_{pq}^{\alpha\alpha}-\gamma_{pq}^{\beta\beta})
\end{gather}
Similarly, $\braket{N^2}$ is
\begin{gather}
\braket{N^2} = 
\sum_{pqrs} \braket{p|q}\braket{r|s} (
-\Gamma_{prsq}^{\alpha\alpha\alpha\alpha}+
\Gamma_{prqs}^{\alpha\beta\alpha\beta}+
\Gamma_{prqs}^{\beta\alpha\beta\alpha}
-\Gamma_{prsq}^{\beta\beta\beta\beta}) \nonumber \\
+ \sum_{ps} \braket{p|s} (\gamma_{ps}^{\alpha\alpha} + \gamma_{ps}^{\beta\beta})
\protect\label{eq:N2}
\end{gather}

\section*{Supplementary Material: Energy expressions} 
\protect\label{sec:energy}
Two-body energy, evaluated from the two-particle density matrix, is
\begin{gather}
\frac{1}{2}\sum_{pqrs}\braket{pq | rs} \Gamma_{\braket{pq|rs}} = 
\frac{1}{2}\sum_{pqrs}\braket{pq | rs} 
\sum_{n} \frac{1}{2n} \frac{1}{\beta}\sum_{i\omega_m} \Tr G \frac{\partial \Sigma^{(n)}}{\partial \lambda}  = \\
\frac{1}{2\beta}\sum_{i\omega_m} \Tr G \Sigma = \braket{V_{ee}}
\end{gather}
The last equality is precisely the Galitskii--Magdal formula.

The spin-integrated energy expression is
\begin{gather}
E = 
\sum_{pq} h_{pq} (\gamma_{pq}^{\alpha\alpha}+\gamma_{pq}^{\beta\beta}) + \\
\frac{1}{2}\sum_{pqrs} v_{\braket{pq | rs}} \left(\Gamma_{\braket{pq|rs}}^{\alpha\alpha\alpha\alpha} +
\Gamma_{\braket{pq|rs}}^{\alpha\beta\alpha\beta} +
\Gamma_{\braket{pq|rs}}^{\beta\alpha\beta\alpha} +
\Gamma_{\braket{pq|rs}}^{\beta\beta\beta\beta}
  \right),
\end{gather}
where all the sums run over AO.

\section*{Supplementary Material: Spin-integrated expressions for two-particle density matrices}
In this section all the lower indices are orbitals. 
The spin label is shown explicitely in the upper indices.
\subsection*{Supplementary Material: Hartree--Fock diagrams}
These are the terms, coming from Hartree--Fock expressions. 
$\gamma$ is the \emph{correlated} one-particle density matrix.
\begin{gather}
\Gamma_{\braket{p_0 q_0 | r_0 s_0}}^{\alpha\alpha\alpha\alpha} 
= \gamma_{p_0 r_0}^{\alpha\alpha} \gamma_{q_0 s_0}^{\alpha\alpha} - \gamma_{p_0 s_0}^{\alpha\alpha}\gamma_{r_0 q_0}^{\alpha\alpha} \protect\label{eq:Gamma_HF_aaaa} \\
\Gamma_{\braket{p_0 q_0| r_0 s_0}}^{\alpha\beta\alpha\beta} = 
 \gamma_{p_0 r_0}^{\alpha\alpha} \gamma_{q_0 s_0}^{\beta\beta} \\ 
\Gamma_{\braket{p_0 q_0 | r_0 s_0}}^{\beta\alpha\beta\alpha} = 
 \gamma_{p_0 r_0}^{\beta\beta} \gamma_{q_0 s_0}^{\alpha\alpha} \\ 
\Gamma_{\braket{p_0 q_0| r_0 s_0}}^{\beta\beta\beta\beta} 
= \gamma_{p_0 r_0}^{\beta\beta} \gamma_{q_0 s_0}^{\beta\beta} 
- \gamma_{p_0 s_0}^{\beta\beta}\gamma_{r_0 q_0}^{\beta\beta} \protect\label{eq:Gamma_HF_bbbb}
\end{gather}

\subsection*{Supplementary Material: GF2 cumulant expressions}
\protect\label{sec:GF2_cum}
\begin{gather}
\Gamma_{\braket{p_0 q_0 |r_0 s_0}}^{\alpha\alpha\alpha\alpha}(GF2) = 
\frac{1}{4}\frac{1}{\beta}\cdot  2\sum_{i\omega_n} 
\sum_{t} ( 
I^{dir,1;\alpha\alpha\alpha\alpha}_{p_0 q_0 t s_0}(i\omega_n) G_{tr_0}^{\alpha\alpha}(i\omega_n) + \nonumber \\
I^{ex,1;\alpha\alpha\alpha\alpha}_{p_0 q_0 t s_0 }(i\omega_n) G_{tr_0}^{\alpha\alpha}(i\omega_n) 
) +   \nonumber \\
\sum_{r} ( 
I^{dir,2;\alpha\alpha\alpha\alpha}_{r q_0 r_0 s_0}(i\omega_n) G_{p_0 r}^{\alpha\alpha}(i\omega_n) + \nonumber \\
I^{ex,2;\alpha\alpha\alpha\alpha}_{r q_0 r_0 s_0}(i\omega_n) G_{p_0 r}^{\alpha\alpha}(i\omega_n)
)
\end{gather}
\begin{gather}
\Gamma_{\braket{p_0 q_0| r_0 s_0}}^{\alpha\beta\alpha\beta}(GF2) = 
\frac{1}{4}\frac{1}{\beta}\cdot  2\sum_{i\omega_n} 
\sum_{t} ( 
I^{dir,1;\alpha\beta\alpha\beta}_{p_0 q_0 t s_0}(i\omega_n) G_{tr_0}^{\alpha\alpha}(i\omega_n) + \nonumber \\
I^{ex,1;\alpha\beta\alpha\beta}_{p_0 q_0 t s_0 }(i\omega_n) G_{tr_0}^{\alpha\alpha}(i\omega_n) 
) +   \nonumber \\
\sum_{r} ( 
I^{dir,2;\alpha\beta\alpha\beta}_{r q_0 r_0 s_0}(i\omega_n) G_{p_0 r}^{\alpha\alpha}(i\omega_n) + \nonumber \\
I^{ex,2;\alpha\beta\alpha\beta}_{r q_0 r_0 s_0}(i\omega_n) G_{p_0 r}^{\alpha\alpha}(i\omega_n)
) = \\
\frac{1}{4}\frac{1}{\beta}\cdot  2\sum_{i\omega_n} 
\sum_{t} ( 
I^{dir,1;\alpha\beta\alpha\beta}_{p_0 q_0 t s_0}(i\omega_n) G_{tr_0}^{\alpha\alpha}(i\omega_n)) + \nonumber \\
\sum_{r} ( 
I^{dir,2;\alpha\beta\alpha\beta}_{r q_0 r_0 s_0}(i\omega_n) G_{p_0 r}^{\alpha\alpha}(i\omega_n)) 
\end{gather}
\begin{gather}
\Gamma_{\braket{p_0 q_0| r_0 s_0}}^{\beta\alpha\beta\alpha}(GF2) = 
\frac{1}{4}\frac{1}{\beta}\cdot  2\sum_{i\omega_n} 
\sum_{t} ( 
I^{dir,1;\beta\alpha\beta\alpha}_{p_0 q_0 t s_0}(i\omega_n) G_{tr_0}^{\beta\beta}(i\omega_n) + \nonumber \\
I^{ex,1;\beta\alpha\beta\alpha}_{p_0 q_0 t s_0 }(i\omega_n) G_{tr_0}^{\beta\beta}(i\omega_n) 
) +   \nonumber \\
\sum_{r} ( 
I^{dir,2;\beta\alpha\beta\alpha}_{r q_0 r_0 s_0}(i\omega_n) G_{p_0 r}^{\beta\beta}(i\omega_n) + \nonumber \\
I^{ex,2;\beta\alpha\beta\alpha}_{r q_0 r_0 s_0}(i\omega_n) G_{p_0 r}^{\beta\beta}(i\omega_n)
) = \\
\frac{1}{4}\frac{1}{\beta}\cdot  2\sum_{i\omega_n} 
\sum_{t} ( 
I^{dir,1;\beta\alpha\beta\alpha}_{p_0 q_0 t s_0}(i\omega_n) G_{tr_0}^{\beta\beta}(i\omega_n)) + \nonumber \\
\sum_{r} ( 
I^{dir,2;\beta\alpha\beta\alpha}_{r q_0 r_0 s_0}(i\omega_n) G_{p_0 r}^{\beta\beta}(i\omega_n)) 
\end{gather}
\begin{gather}
\Gamma_{\braket{p_0 q_0 |r_0 s_0}}^{\beta\beta\beta\beta}(GF2) = 
\frac{1}{4}\frac{1}{\beta}\cdot  2\sum_{i\omega_n} 
\sum_{t} ( 
I^{dir,1;\beta\beta\beta\beta}_{p_0 q_0 t s_0}(i\omega_n) G_{tr_0}^{\beta\beta}(i\omega_n) + \nonumber \\
I^{ex,1;\beta\beta\beta\beta}_{p_0 q_0 t s_0 }(i\omega_n) G_{tr_0}^{\beta\beta}(i\omega_n) 
) +   \nonumber \\
\sum_{r} ( 
I^{dir,2;\beta\beta\beta\beta}_{r q_0 r_0 s_0}(i\omega_n) G_{p_0 r}^{\beta\beta}(i\omega_n) + \nonumber \\
I^{ex,2;\beta\beta\beta\beta}_{r q_0 r_0 s_0}(i\omega_n) G_{p_0 r}^{\beta\beta}(i\omega_n)
)
\end{gather}
\begin{gather}
I^{dir,1;\alpha\alpha\alpha\alpha}_{p_0 q_0 t s_0}(\tau) = 
- \sum_{uvw} v_{\braket{tu | vw}} G_{vp_0}^{\alpha\alpha}(\tau) G_{wq_0}^{\alpha\alpha} (\tau) 
G_{s_0u}^{\alpha\alpha} (-\tau)  \\
I^{dir,2;\alpha\alpha\alpha\alpha}_{r q_0 r_0 s_0}(\tau) = 
- \sum_{pqs} v_{\braket{pq | rs}} G_{r_0 p}^{\alpha\alpha}(\tau) G_{s_0 q}^{\alpha\alpha} (\tau) 
G_{s q_0}^{\alpha\alpha} (-\tau) \\
I^{ex,1;\alpha\alpha\alpha\alpha}_{p_0 q_0 t s_0 }(\tau) = 
\sum_{uvw} v_{\braket{tu | vw}} 
G_{wp_0}^{\alpha\alpha}(\tau) G_{vq_0}^{\alpha\alpha} (\tau) G_{s_0u}^{\alpha\alpha} (-\tau)  \\
I^{ex, 2;\alpha\alpha\alpha\alpha}_{r q_0 r_0 s_0}(\tau) = 
\sum_{pqs} v_{\braket{pq | rs}}
G_{s_0 p}^{\alpha\alpha}(\tau) G_{r_0 q}^{\alpha\alpha} (\tau) G_{sq_0}^{\alpha\alpha} (-\tau)  
\end{gather}
\begin{gather}
I^{dir,1;\alpha\beta\alpha\beta}_{p_0 q_0 t s_0}(\tau) = 
- \sum_{uvw} v_{\braket{tu | vw}} G_{vp_0}^{\alpha\alpha}(\tau) G_{wq_0}^{\beta\beta} (\tau) 
G_{s_0u}^{\beta\beta} (-\tau)  \\
I^{dir,2;\alpha\beta\alpha\beta}_{r q_0 r_0 s_0}(\tau) = 
- \sum_{pqs} v_{\braket{pq | rs}} G_{r_0 p}^{\alpha\alpha}(\tau) G_{s_0 q}^{\beta\beta} (\tau) 
G_{s q_0}^{\beta\beta} (-\tau) \\
I^{ex,1;\alpha\beta\alpha\beta}_{p_0 q_0 t s_0 }(\tau) = 0 
\text{ due to zero integrals }\braket{\alpha\beta|\beta\alpha} \\
I^{ex, 2;\alpha\beta\alpha\beta}_{r q_0 r_0 s_0}(\tau) = 0 \text{ due to zero integrals }\braket{\beta\alpha|\alpha\beta} 
\end{gather}
\begin{gather}
I^{dir,1;\beta\alpha\beta\alpha}_{p_0 q_0 t s_0}(\tau) = 
- \sum_{uvw} v_{\braket{tu | vw}} G_{vp_0}^{\beta\beta}(\tau) G_{wq_0}^{\alpha\alpha} (\tau) 
G_{s_0u}^{\alpha\alpha} (-\tau)  \\
I^{dir,2;\beta\alpha\beta\alpha}_{r q_0 r_0 s_0}(\tau) = 
- \sum_{pqs} v_{\braket{pq | rs}} G_{r_0 p}^{\beta\beta}(\tau) G_{s_0 q}^{\alpha\alpha} (\tau) 
G_{s q_0}^{\alpha\alpha} (-\tau) \\
I^{ex,1;\beta\alpha\beta\alpha}_{p_0 q_0 t s_0 }(\tau) = 0 \\
I^{ex, 2;\beta\alpha\beta\alpha}_{r q_0 r_0 s_0}(\tau) =  0 
\end{gather}
Equations for $I^{\beta\beta\beta\beta}$ are the same as for $I^{\alpha\alpha\alpha\alpha}$ with full $\alpha \rightarrow \beta$ replacement.

\subsection*{Supplementary Material: GW cumulant}
Lack of correlated exchange makes the spin-integrated expression very simple:
\begin{gather}
\Gamma_{(p_0 q_0 | r_0 s_0)}^{\alpha\alpha\alpha\alpha} = 
\frac{1}{\beta}\sum_{\Omega_m} 
\sum_{pqrs} \Pi_{r_0 s_0 pq}^{\alpha\alpha\alpha\alpha}(\Omega_m) W_{(pq|rs)}^{\alpha\alpha\alpha\alpha}(\Omega_m) \Pi_{rs p_0 q_0}^{\alpha\alpha\alpha\alpha}(\Omega_m) = \\
\frac{1}{\beta}\sum_{\Omega_m} 
\sum_{pqrs} \Pi_{r_0 s_0 pq}^{\alpha\alpha\alpha\alpha}(\Omega_m) V_{pq}^Q (\delta_{Q,Q^\prime}+\tilde{P}_{QQ^\prime}(\Omega_m))  V^{Q^\prime}_{rs}\Pi_{rs p_0 q_0}^{\alpha\alpha\alpha\alpha}(\Omega_m)  \\
\Gamma_{(p_0 q_0 | r_0 s_0)}^{\alpha\alpha\beta\beta} = 
\frac{1}{\beta}\sum_{\Omega_m} 
\sum_{pqrs} \Pi_{r_0 s_0 pq}^{\beta\beta\beta\beta}(\Omega_m) W_{(pq|rs)}^{\beta\beta\alpha\alpha}(\Omega_m) \Pi_{rs p_0 q_0}^{\alpha\alpha\alpha\alpha}(\Omega_m) = \\
\frac{1}{\beta}\sum_{\Omega_m} 
\sum_{pqrs} \Pi_{r_0 s_0 pq}^{\beta\beta\beta\beta}(\Omega_m) V_{pq}^Q (\delta_{Q,Q^\prime}+\tilde{P}_{QQ^\prime}(\Omega_m))  V^{Q^\prime}_{rs}\Pi_{rs p_0 q_0}^{\alpha\alpha\alpha\alpha}(\Omega_m)  \\
\end{gather}

\section*{Supplementary Material: Finite-difference calculations}
\begin{table} [tbh!]
  \caption{Convergence of central $n$-point stencil numerical differentiation of $\braket{N}$ with respect to $\mu$ for Be atom, \textbf{solution II} at $\beta = 1000$~a.u.$^{-1}$. 
  The chemical potential step size is $1$~mHartree.
\protect\label{tbl:fdiff_n_mu}}
\begin{tabular}{l|c}
\hline
Stencil    &  $\partial \braket{N}/\partial \mu$, a.u.$^{-1}$ \\
\hline
3-point    & 3.47121523 \\
5-point    & 3.47122213 \\
7-point    & 3.47122591 
\end{tabular}
\end{table}

\section*{Supplementary Material: Finite-temperature graphs}
\protect\label{sec:finite_T_graphs}
\begin{figure}[!h]
  \includegraphics[width=8cm]{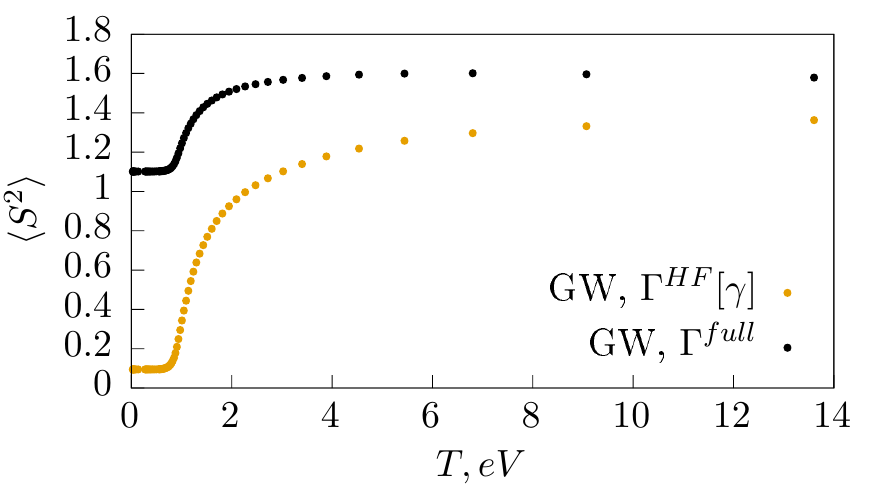}
  \includegraphics[width=8cm]{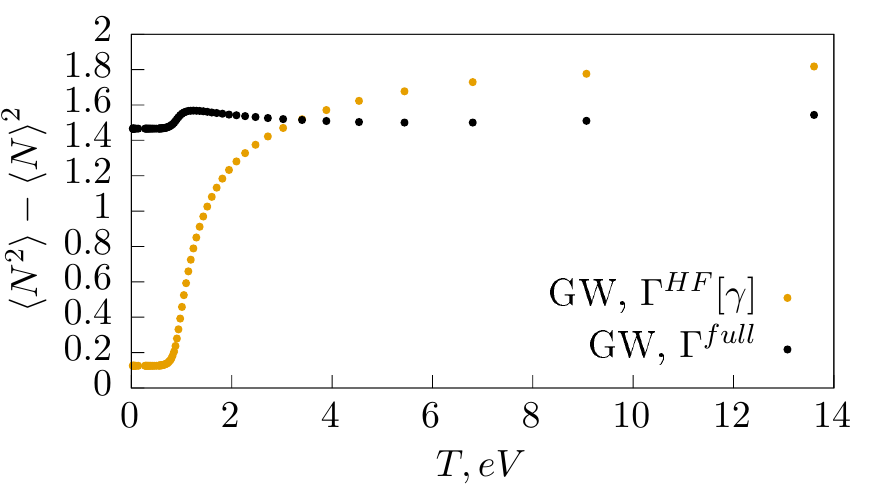} \\
  \includegraphics[width=8cm]{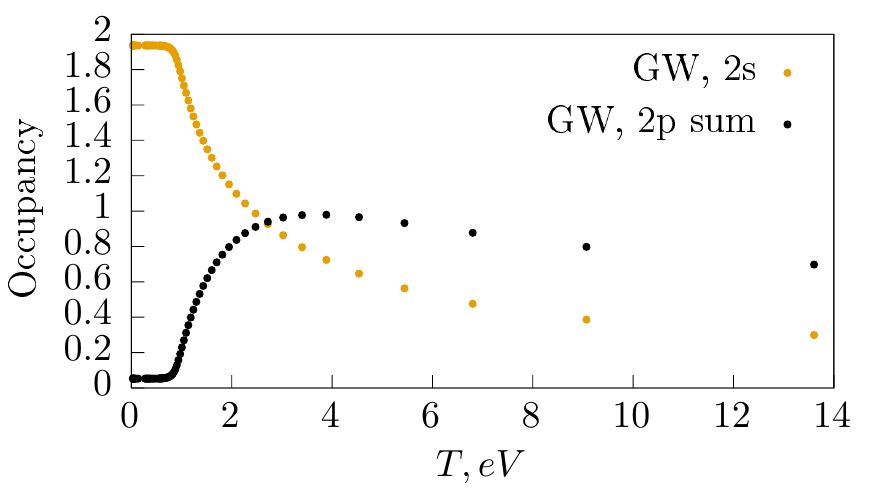}
\centering
\caption{Top: Temperature dependence of $\braket{S^2}$ and $\braket{N^2}-\braket{N}^2$ for a Be atom, 
         computed with GW. 
         Contributions of antisymmetrized Kroneker product of one-particle density matrices  
         and full two-particle contributions are shown.
         Bottom: 2s and 2p occupancies at different temperatures.
         \protect\label{fig:Be_GW}}
\end{figure}

\begin{figure}[!h]
  \includegraphics[width=8cm]{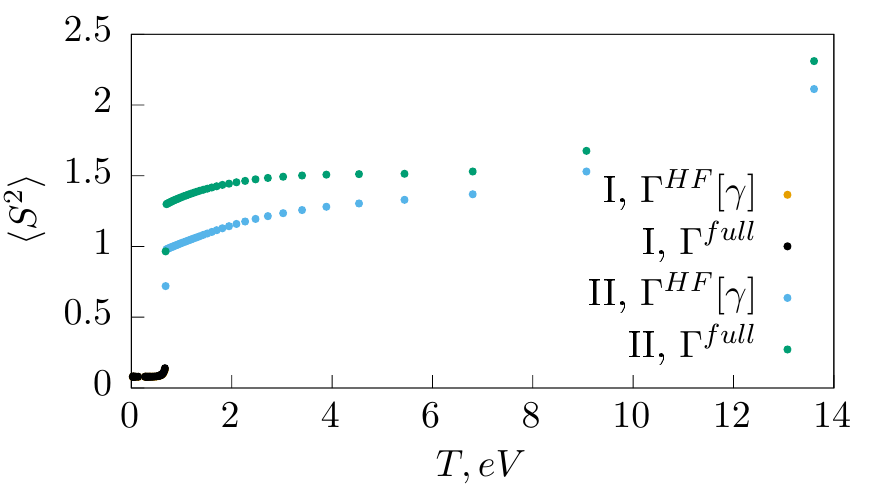}
  \includegraphics[width=8cm]{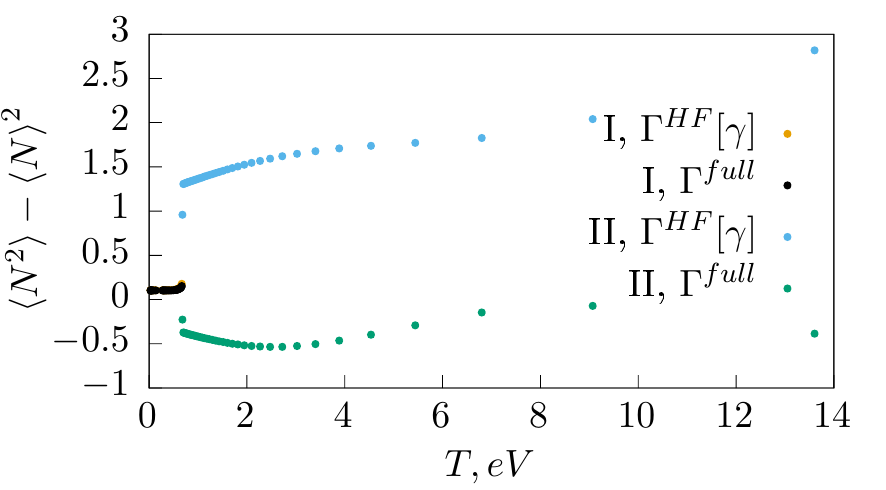} \\
  \includegraphics[width=8cm]{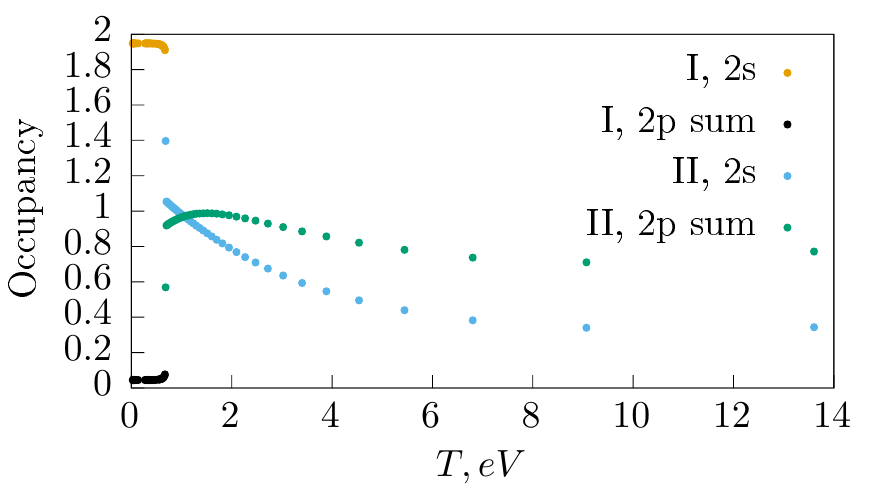}
\centering
\caption{Top: Temperature dependence of $\braket{S^2}$ and $\braket{N^2}-\braket{N}^2$ for a Mg atom, 
         computed with GF2. 
         Contributions of antisymmetrized Kroneker product of one-particle density matrices  
         and full two-particle contributions are shown.
         The \textbf{solution I} corresponds to a closed shell at the low temperature. 
         The \textbf{solution II} is the high-temperature solution with nearly equal occupations of 2s and 2p orbitals. 
         Its low-temperature limit corresponds to a 
         broken-spin Hartree--Fock solution with fractional occupations. 
         Bottom: 2s and 2p occupancies at different temperatures.
         \protect\label{fig:Mg_GF2}}
\end{figure}

\begin{figure}[!h]
  \includegraphics[width=8cm]{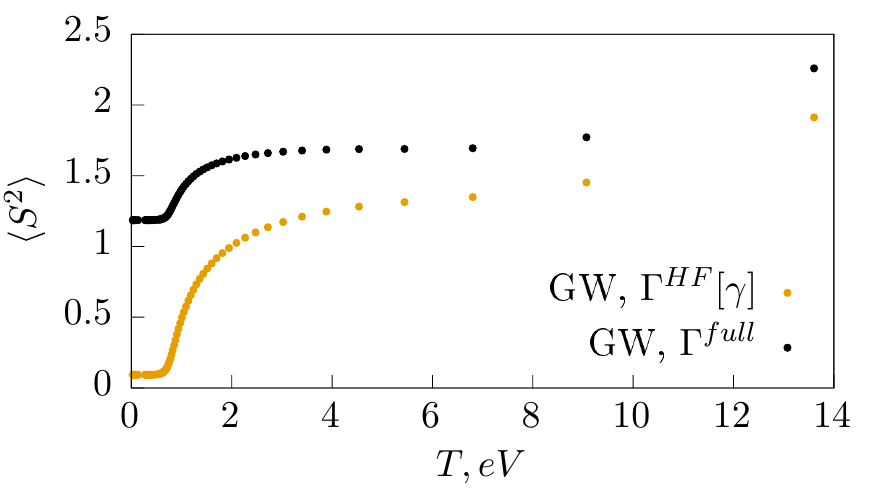}
  \includegraphics[width=8cm]{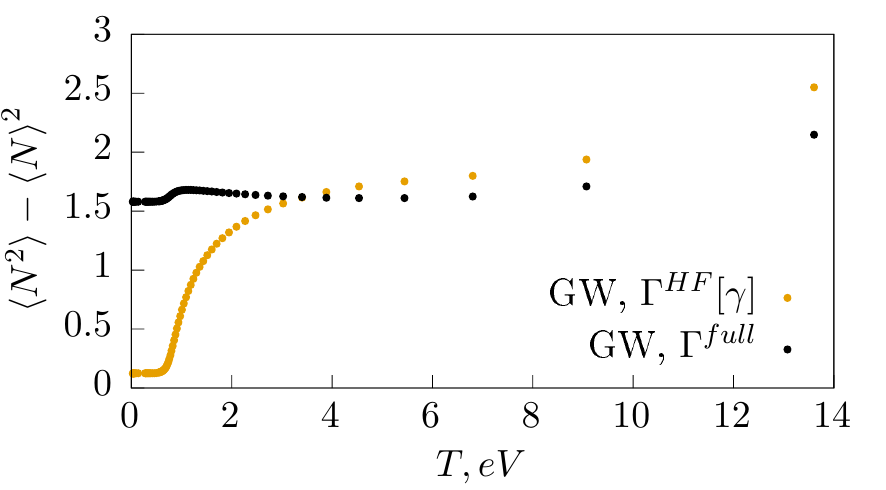} \\
  \includegraphics[width=8cm]{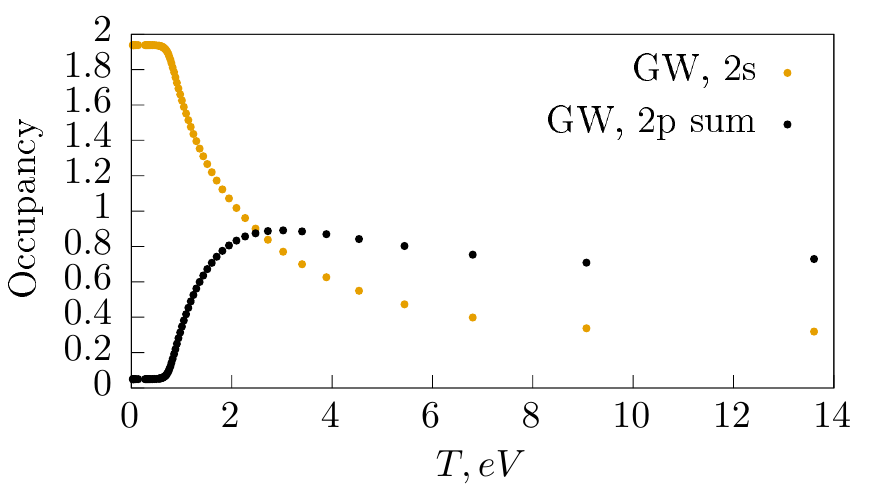}
\centering
\caption{Top: Temperature dependence of $\braket{S^2}$ and $\braket{N^2}-\braket{N}^2$ for a Mg atom, 
         computed with GW. 
         Contributions of antisymmetrized Kroneker product of one-particle density matrices  
         and full two-particle contributions are shown.
         Bottom: 2s and 2p occupancies at different temperatures.
         \protect\label{fig:Mg_GW}}
\end{figure}

\begin{figure}[!h]
  \includegraphics[width=8cm]{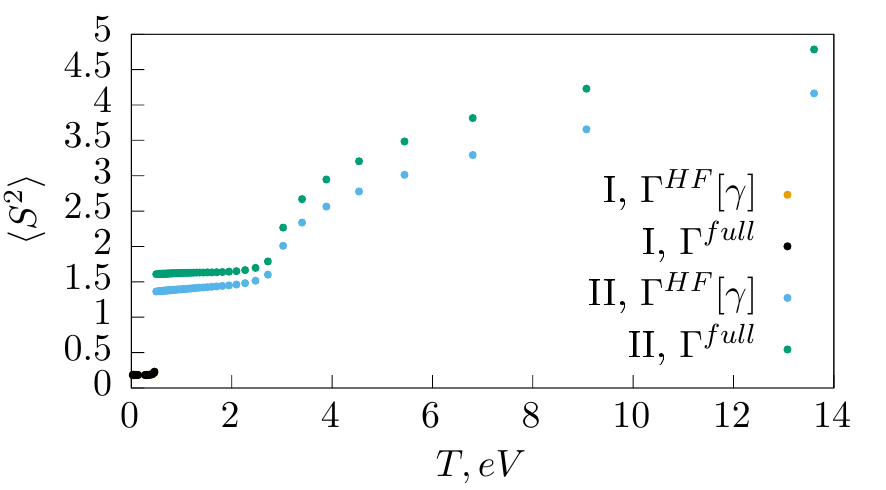}
  \includegraphics[width=8cm]{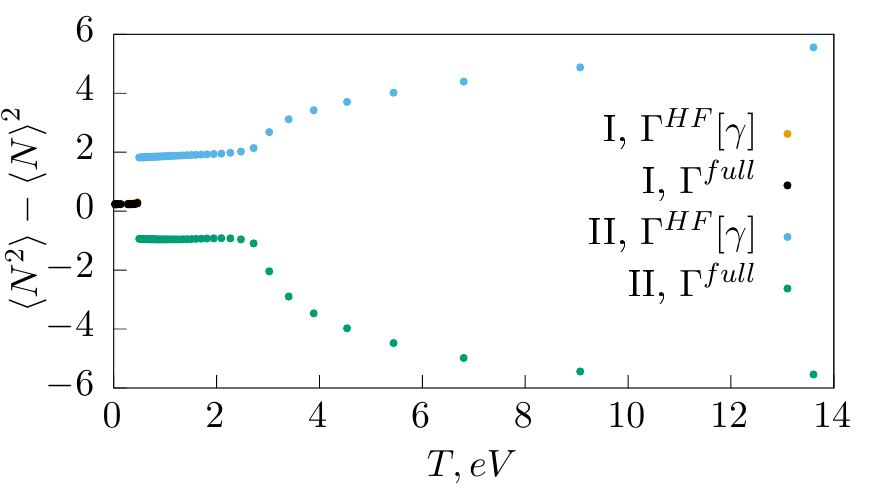} \\
  \includegraphics[width=8cm]{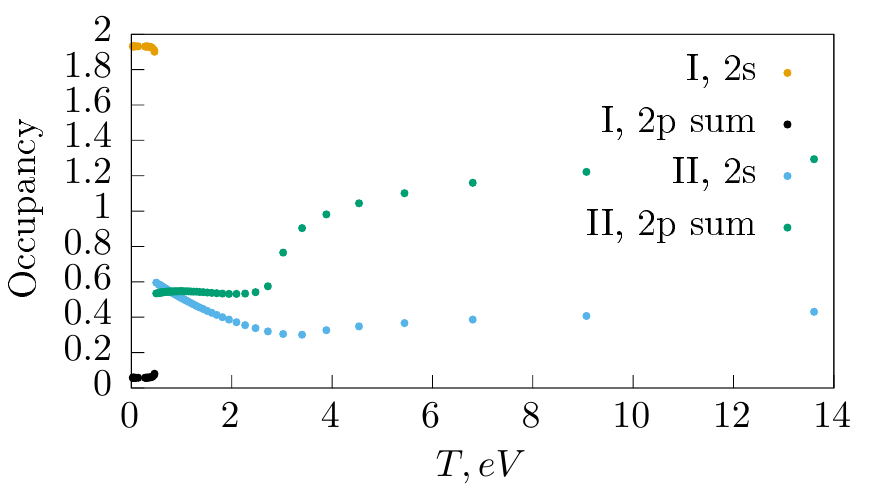}
\centering
\caption{Top: Temperature dependence of $\braket{S^2}$ and $\braket{N^2}-\braket{N}^2$ for a Ca atom, 
         computed with GF2. 
         Contributions of antisymmetrized Kroneker product of one-particle density matrices  
         and full two-particle contributions are shown.
         The \textbf{solution I} corresponds to a closed shell at the low temperature. 
         The \textbf{solution II} is the high-temperature solution with nearly equal occupations of 2s and 2p orbitals. 
         Its low-temperature limit corresponds to a 
         broken-spin Hartree--Fock solution with fractional occupations. 
         Bottom: 2s and 2p occupancies at different temperatures.
         \protect\label{fig:Ca_GF2}}
\end{figure}

\begin{figure}[!h]
  \includegraphics[width=8cm]{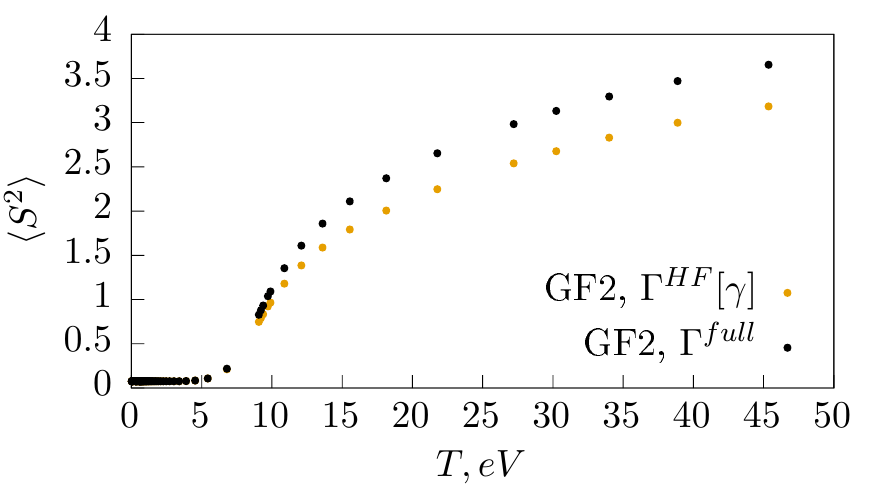}
  \includegraphics[width=8cm]{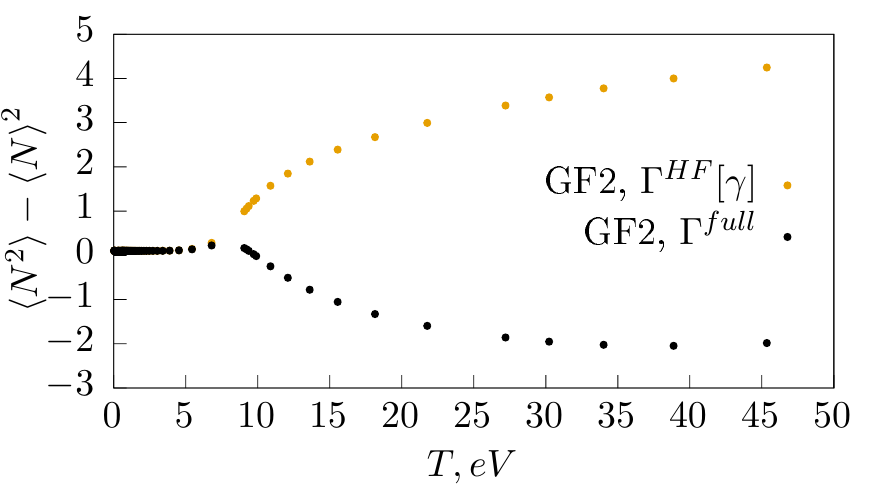} 
\centering
\caption{Temperature dependence of $\braket{S^2}$ and $\braket{N^2}-\braket{N}^2$ for a Ne atom, 
         computed with GF2. 
         Contributions of antisymmetrized Kroneker product of one-particle density matrices  
         and full two-particle contributions are shown.
         \protect\label{fig:Ne_GF2}}
\end{figure}

\begin{figure}[!h]
  \includegraphics[width=8cm]{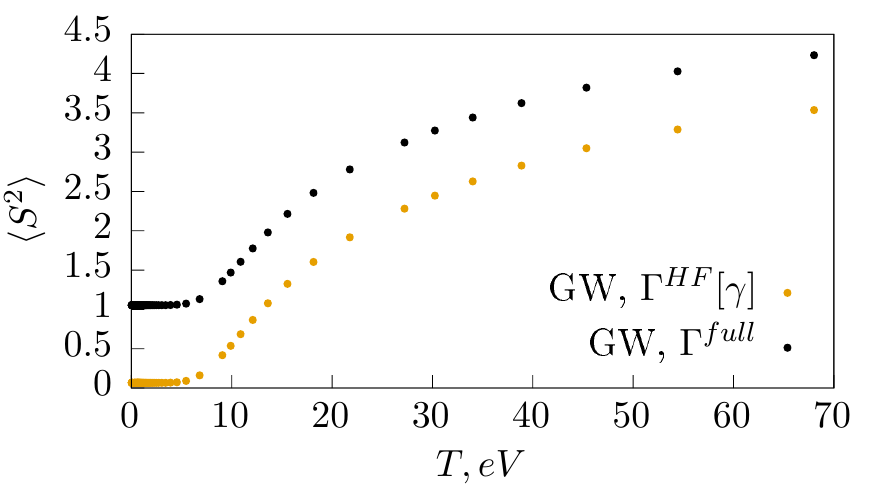}
  \includegraphics[width=8cm]{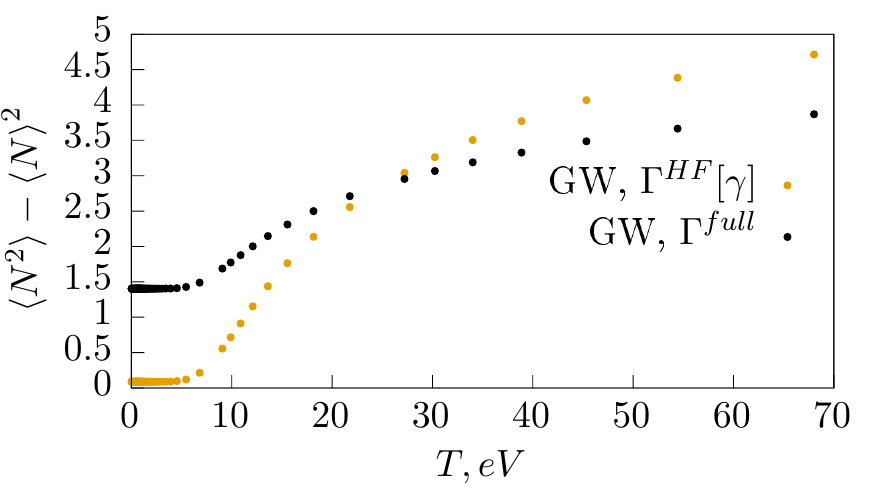} 
\centering
\caption{Temperature dependence of $\braket{S^2}$ and $\braket{N^2}-\braket{N}^2$ for a Ne atom, 
         computed with GW. 
         Contributions of antisymmetrized Kroneker product of one-particle density matrices  
         and full two-particle contributions are shown.
         \protect\label{fig:Ne_GW}}
\end{figure}

\begin{figure}[!h]
  \includegraphics[width=8cm]{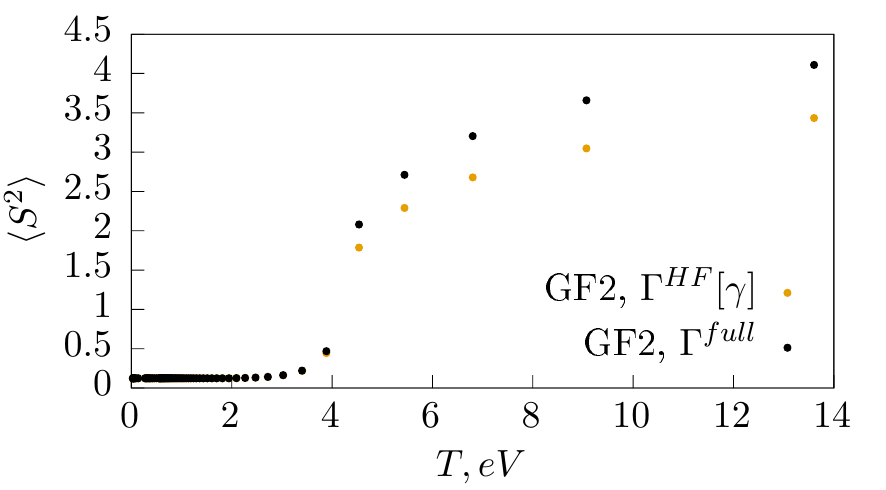}
  \includegraphics[width=8cm]{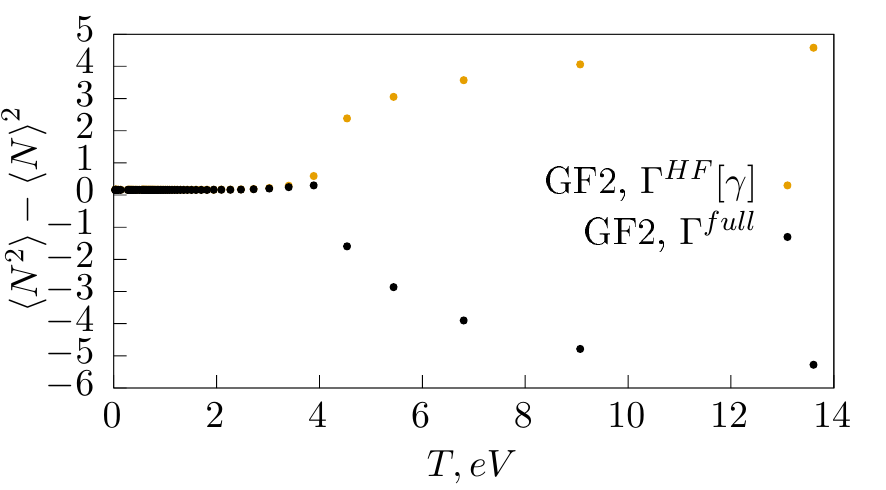} 
\centering
\caption{Temperature dependence of $\braket{S^2}$ and $\braket{N^2}-\braket{N}^2$ for a Ar atom, 
         computed with GF2. 
         Contributions of antisymmetrized Kroneker product of one-particle density matrices  
         and full two-particle contributions are shown.
         \protect\label{fig:Ar_GF2}}
\end{figure}

\begin{figure}[!h]
  \includegraphics[width=8cm]{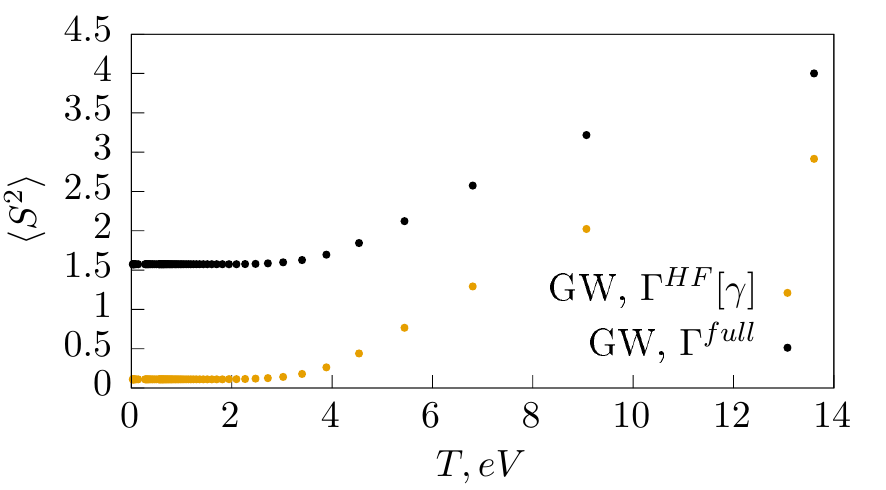}
  \includegraphics[width=8cm]{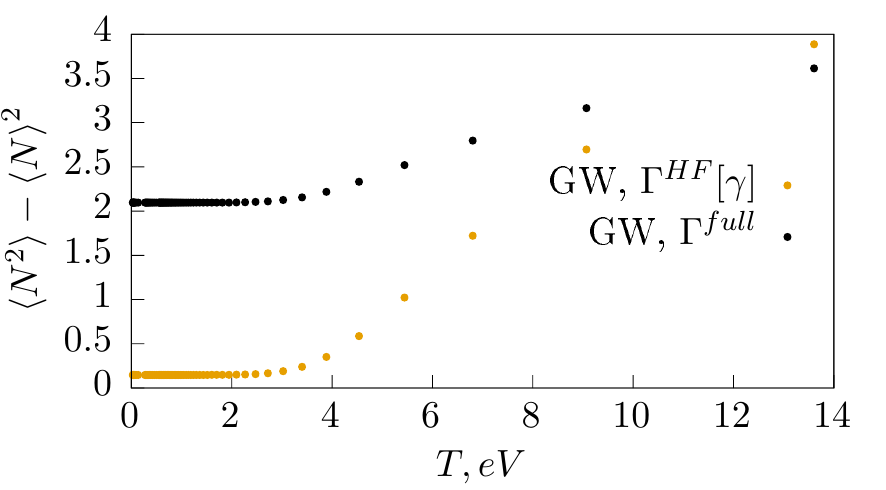} 
\centering
\caption{Temperature dependence of $\braket{S^2}$ and $\braket{N^2}-\braket{N}^2$ for a Ar atom, 
         computed with GW. 
         Contributions of antisymmetrized Kroneker product of one-particle density matrices  
         and full two-particle contributions are shown.
         \protect\label{fig:Ar_GW}}
\end{figure}

\section*{Appendix A: Independent electrons}
Consider a one-electron perturbation $\hat{O}$ of the Hamiltonian of a system of independent electrons:
\begin{gather}
H_0(\lambda) = H_0 + \lambda O,
\end{gather}
such that the atomic orbitals (AO) are not perturbed. 
The inverse Green's function in AO is
\begin{gather}
(\mathbf{G}_0)^{-1}(i\omega_n) = i\omega_n \mathbf{S} +\mu \mathbf{N} - \mathbf{H}_0, 
\protect\label{eq:G_0_matrix}
\end{gather}
where the bold font denotes matrices in the AO basis and $\mathbf{S}$ is the AO overlap matrix.
Its derivative returns the perturbation operator in the AO basis
\begin{gather}
\frac{d}{d \lambda} (\mathbf{G}_0)^{-1}(i\omega_n) = -\mathbf{O}.
\protect\label{eq:der_g0_inv}
\end{gather}
The grand potential of a system of independent particles is defined as 
\begin{gather}
\Omega_0 = -\frac{1}{\beta} \sum_{\omega_n} \Tr \ln (-G_0^{-1} (i\omega_n) ).
\end{gather}
The derivative of the grand potential is 
\begin{gather}
\frac{d}{d \lambda}\Omega = -\frac{1}{\beta} \sum_{\omega_n} \Tr \frac{ (\frac{d}{d\lambda}G_0^{-1} (i\omega_n) ) }{(G_0^{-1} (i\omega_n) )}  = 
\frac{1}{\beta} \sum_{\omega_n} \Tr O G_0(i\omega_n). 
\end{gather}
Thus, the derivative approach and the expectation value approach are yielding equivalent results for independent electrons. 

\subsection*{Appendix B: Constant $\braket{N}$}
Often it is desirable to preserve the average number of electrons. 
The properties in this case can be found from 
the derivative of the Helmholtz free energy with the optimized $\mu$:
\begin{gather}
F = \Omega[G] + \mu \braket{N} = \Phi[G]  -\frac{1}{\beta} \sum_{\omega_n} \Tr \Sigma G - \frac{1}{\beta}\sum_{\omega_n}\Tr\ln(-G^{-1})  + \mu \braket{N}.
\end{gather}
To derive the analytical expression, we can consider a perturbation of 
both chemical potential $\mu(\lambda)$ and the one-electron Hamiltonian.
Consequently, the derivative of the  the Helmholtz free energy with respect to $\lambda$ is 
\begin{gather}
\frac{dF}{d\lambda} = 
\left( \frac{\partial F}{\partial \lambda} \right)_{G, \braket{N}} = 
-\frac{1}{\beta}\sum_{\omega_n} \Tr \left(\frac{\partial\Sigma}{\partial \lambda}
\right)_{G,v} G
+ \frac{\partial \mu}{\partial \lambda}  \braket{N}.
\end{gather}
Because of the dependence of $\mu$ on $\lambda$, 
the expression for the derivative of $G_0^{-1}$ changes to
\begin{gather}
\frac{d \mathbf{G}_0^{-1}}{d \lambda} = \frac{\partial \mu}{\partial \lambda} \mathbf{N} - \mathbf{O}.
\end{gather}
The free energy derivative then becomes
\begin{gather}
\left( \frac{\partial F}{\partial \lambda} \right)_{G,\braket{N}} = 
-\frac{1}{\beta} \sum_{\omega_n} \Tr \big[ \frac{\partial \mu}{\partial \lambda} NG - O G \big] 
+ \frac{\partial \mu}{\partial \lambda}  \braket{N} = \\
\braket{O} 
- \frac{\partial \mu}{\partial \lambda}  \braket{N} 
+ \frac{\partial \mu}{\partial \lambda}  \braket{N} = \braket{O} = 
\left( \frac{\partial \Omega}{\partial \lambda} \right)_{G,\mu}[\mu=\mu(\braket{N})]. 
\end{gather}
Thus, differentiation of $F$ gives the same result as differentiation of $\Omega$ at the same chemical potential. 
The proof for two-particle properties is analogous to the proof for one-particle perturbations.

\section*{Appendix C: Numerical algorithm for finding GF2 density cumulants}
The intermediates used in GF2 have 4 orbital indices and one time index making them very bulky objects when memory demands are concerned.
However, two-particle density cumulant does not require storage of the intermediate at all time or frequency points. 
\begin{enumerate}
\item Find $G$ solving the GF2 Dyson equation to self-consistency.
\item Compute a special transformation matrix
\begin{gather}
T^{\text{sp}}(i\omega, \tau) = T(\tau=0, i\omega) \cdot T(i\omega, \tau),
\end{gather}
where $T(\tau, \omega)$ transforms from the Matsubara frequency to the imaginary time (through the intermediate representation),
$T(\omega, \tau)$ transforms from the imaginary time to the Matsubara frequency. 
The $T(\tau=0, i\omega)$ multiplier gives a weight of each frequency for the non-uniform grid 
when the sums over all frequencies is performed.
\item Evaluate 
\begin{gather}
\tilde{G}(\tau) = \sum_\omega (T^{\text{sp}}(i\omega, \tau))^\dagger T(i\omega, \tau) G(\tau).
\end{gather}
This transformed Green's function is needed to move the summation over frequencies to the imaginary time
\begin{gather}
\sum_\omega \text{weight}(i\omega) G(i\omega) I(i\omega) = 
\sum_{\omega,\tau} \text{weight}(i\omega) G(i\omega) T(i\omega, \tau) I(\tau) = \\
\sum_{\omega,\tau} (T(i\omega, \tau))^\dagger \text{weight}(i\omega) G(i\omega)  I(\tau) = 
\sum_{\tau} \tilde{G}(\tau)  I(\tau).  
\end{gather}
\item For each $\tau$, evaluate intermediates with Eqs.~\ref{eq:Idir1}--\ref{eq:Iex2} 
($N^5$ implementation contracts each integral index sequentially as done in AO-to-MO integral transformations) 
and absorb it into a cumulant 
\begin{gather}
\Gamma_{\braket{p_0 q_0 | r_0 s_0}}^\text{GF2} +=
\sum_{t} ( 
I^{dir,1}_{p_0 q_0 t s_0}(\tau) \tilde{G}_{tr_0}(\tau) + 
I^{ex,1}_{p_0 q_0 t s_0 }(\tau) \tilde{G}_{tr_0}(\tau) ). 
\end{gather}
\end{enumerate}
Thus, only the storage at the running $\tau$ is needed. 
Again, note  that the contractions are similar to the ones in the AO to MO transformation of integrals. 
The RI approximation allows to perform these operations efficiently, 
contracting RI 3-index tensors with four Green's functions first 
and contracting over the auxiliary basis index at the last step. 

\section*{Appendix D: Numerical algorithm for finding GW density cumulants}
The GW two-particle density cumulant (Eq.~\ref{eq:GW_cumulant}) is different from the GF2 one due to the 
frequency-dependent polarization function. 
This complicates the summation over frequencies, and the trick that was used for GF2 cannot be applied directly. 
Our implementation relies on evaluation of contractions between 3-index RI integrals and the polarization function $\Pi$. 
We define this intermediate as
\begin{gather}
I^{Q^\prime}_{p_0 q_0}(\Omega_m) = \sum_{rs} V^{Q^\prime}_{rs}\Pi_{rs p_0 q_0}(\Omega_m).
\end{gather}
However, the full polarization function $\Pi$ is a very large quantity. 
Rather than computing it directly, 
we contract each integral index sequentially with the Green's function in the time domain for each $\tau$ point
\begin{gather}
I1^{Q^\prime}_{q_0 s}(\tau) = \sum_{r} V^{Q^\prime}_{rs} G_{q_0 r}(\tau), \\
I^{Q^\prime}_{p_0 q_0}(\tau) = \sum_{s} I1^{Q^\prime}_{q_0 s} G_{s p_0}(-\tau). 
\end{gather}
Although it is possible to evaluate $I^Q_{pq}(\Omega)$ on the fly computating $I^Q_{pq}(\tau)$, 
transforming it to the frequency domain (``direct'' algorithm), 
and absorbing into the two-particle density matrix, 
this scheme does not scale well with respect to the grid size despite its low storage requirements, 
increasing computational expense due to re-evaluations of the intermediates.  
A more efficient implementation (``semi-direct'' algorithm) computes the entire intermediate first and stores it on disk:

For each $\tau$:
\begin{enumerate}
\item Evaluate $I^{Q^\prime}_{pq}$ at a given time point $\tau$.
\item Write $I^{Q^\prime}_{pq}$ at this time point $\tau$ on disk.
\end{enumerate}
When the intermediate is computed for all time points, 
perform the transformation to the frequency domain:

For each bosonic frequency $\Omega$:
\begin{enumerate}
\item Set temporary work space to zero---the 3-index tensor $J^{Q^\prime}_{pq}$.
\item For each $\tau$:
    \begin{enumerate}
        \item Read $I^{Q^\prime}_{pq}$ for a given time point $\tau$ from the disk.
        \item Accumulate into $J += T(\Omega,\tau)\cdot I$
    \end{enumerate}
\item Write $J$ at a frequency point $\Omega$ on disk.
\end{enumerate}
This procedure does not require keeping the intermediate at the entire time/frequency domain in memory---instead, 
it uses disk to store the intermediate and keeps only the running time/frequency slice of the intemediate in memory. 
Since there are only two non-zero spin blocks of $\Pi$ ($\alpha\alpha\alpha\alpha$ and $\beta\beta\beta\beta$), 
only two spin blocks of intemediates are computed. 
Finally, the two-particle density matrix cumulant is assembled:

For each bosonic frequency $\Omega$:
\begin{enumerate}
\item Read $I^{Q^\prime}_{pq}$ at a frequency point $\Omega$ from the disk.
\item Evaluate and accumulate $\Gamma_{(rs|pq)} += weight(\Omega) \cdot I_{pq}^Q(\delta_{Q,Q^\prime}+\tilde{P}_{QQ^\prime}(\Omega_m))I^{Q^\prime}_{rs})$.
\end{enumerate}
The expression above is implemented through matrix multiplications as $weight \cdot \mathbf{I}^T (\mathbf{1} + \tilde{\mathbf{P}}) \mathbf{I}$. 
The weight here has the same origin as in GF2---it connects frequency representation 
with the time representation at a zero time point, serving as a weight for a non-uniform grid.

\clearpage
\bibliographystyle{prf}

\begin{thebibliography}{100}
\bibitem{note:CPSCF}
When an ensemble is described approximately, the higher-order grand-potential
  derivatives are not necessarily equal to the expectation values of a
  higher-order excitation operator. For example, the second derivative of the
  Hartree--Fock energy leads to well-known coupled-perturbed self-consistent
  field (CPSCF) equations\cite{Pulay:CPHF:69} which are usually solved
  iteratively.
\bibitem{note:coulomb}
Sometimes in physical literature the electron repulsion integrals (ERI) are
  named ``Coulomb integrals''. The latter term can be confused with the Coulomb
  part of the Fock matrix in Hartree--Fock approximation. ERI, or just
  ``two-electron integrals'', widely accepted in chemical literature, avoid
  this ambiguity.
\bibitem{note:ExpVal}
Expectation values (in a strong sense) are defined only if the density operator
  $\hat{\rho}$ is defined. Approximate Green's functions may not be ensemble
  representable; in such a case the density operator and expectation-value
  definitions do not exist. However, one can define a value of a property
  through summation over Matsubara frequencies and trace with the Green's
  function. The latter definition is an expectation value in a ``weak'' sense,
  because the ensemble density operator does not have to exist.
\bibitem{note:crossing}
In the Green's function literature antisymmetric structure of Green's functions
  and vertex functions is sometimed called a ``crossing symmetry''.
\end{thebibliography}

\end{document}